\documentclass[aps,prd,nofootinbib,superscriptaddress]{revtex4}

\usepackage{epsfig}
\input psfig.sty
\usepackage{graphicx}
\usepackage{dcolumn}
\usepackage{bm}
\usepackage{amsmath} 
\usepackage[english]{babel}
\usepackage{mathrsfs}
\usepackage{amsfonts}
\usepackage{amssymb}
\usepackage{hyperref}
\usepackage{longtable}
\usepackage{float}
\usepackage{dcolumn}
\usepackage{appendix}
\usepackage{multirow}
\usepackage{longtable}
\usepackage{color}
\usepackage{makecell}

\newcommand{\mytilde}{\raise.17ex\hbox{$\scriptstyle\mathtt{\sim}$}}

\newcommand{\bea}{\begin{eqnarray*}}
\newcommand{\eea}{\end{eqnarray*}}

\newcommand{\tc}{\eta_{{k}}^c}
\newcommand{\nn}{\nonumber \\}
\newcommand{\zk}{|z_k|}
\newcommand{\mH}{\mathcal{H}}
\newcommand{\mP}{\mathcal{P}}
\newcommand{\mA}{\mathcal{A}}
\newcommand{\dphi}{\delta \phi}
\newcommand{\x}{\textbf{x}}

\newcommand{\bra}{\langle}
\newcommand{\ket}{\rangle}
\newcommand{\nk}{\textbf{k}}
\newcommand{\ann}{\hat{a}} 
\newcommand{\cre}{\hat{a}^{\dagger}}    

\newcommand{\wig}{\textit{Wigner's}} 

  \newcommand{\DN}{  \Delta N_*}  
  \newcommand{\std}{\textrm{std}}  
  \newcommand{\fin}{\textrm{end}}
\newcommand{\ini}{\textrm{ini}}    
\newcommand{\reh}{\textrm{reh}}    
\newcommand{\nuc}{\textrm{nuc}}    
\newcommand{\rad}{\textrm{rad}}    

\newcommand{\ei}{\epsilon_1}
\newcommand{\eii}{\epsilon_2}
\newcommand{\mI}{\mathcal{I}}

\begin{document}

\title{Observational constraints on inflationary potentials within the quantum collapse framework}

\author{Gabriel Le\'{o}n}
\email{gleon@fcaglp.unlp.edu.ar }
\affiliation{Grupo de Astrof\'{\i}sica, Relatividad y Cosmolog\'{\i}a, Facultad 
	de Ciencias Astron\'{o}micas y Geof\'{\i}sicas, Universidad Nacional de La 
	Plata, Paseo del Bosque S/N 1900 La Plata, Argentina.\\
	CONICET, Godoy Cruz 2290, 1425 Ciudad Aut\'onoma de Buenos Aires, Argentina. }

\author{Alejandro Pujol}
\email{apujol@df.uba.ar}
\affiliation{Departamento de F\'isica and IFIBA, Facultad de Ciencias Exactas y Naturales, Universidad de Buenos Aires, Ciudad Universitaria - Pab. I, Buenos Aires 1428, Argentina.\\
	CONICET, Godoy Cruz 2290, 1425 Ciudad Aut\'onoma de Buenos Aires, Argentina.}

\author{Susana J. Landau}
\email{slandau@df.uba.ar }
\affiliation{Departamento de F\'isica and IFIBA, Facultad de Ciencias Exactas y Naturales, Universidad de Buenos Aires, Ciudad Universitaria - Pab. I, Buenos Aires 1428, Argentina.\\
	CONICET, Godoy Cruz 2290, 1425 Ciudad Aut\'onoma de Buenos Aires, Argentina.}

\author{Mar\'{\i}a P\'{\i}a Piccirilli}
\email{mpp@fcaglp.unlp.edu.ar }
\affiliation{Grupo de Astrof\'{\i}sica, Relatividad y Cosmolog\'{\i}a, Facultad 
	de Ciencias Astron\'{o}micas y Geof\'{\i}sicas, Universidad Nacional de La 
	Plata, Paseo del Bosque S/N 1900 La Plata, Argentina.\\
	CONICET, Godoy Cruz 2290, 1425 Ciudad Aut\'onoma de Buenos Aires, Argentina. }

\begin{abstract}

The physical mechanism responsible for the emergence of primordial cosmic seeds from a perfect isotropic and homogeneous Universe has not been fully 
addressed in standard cosmic inflation. To handle this  shortcoming, D. Sudarsky et al have developed a proposal: the self-induced collapse hypothesis. In this scheme, the objective collapse of the inflaton's wave function generates the inhomogeneity and anisotropy at all scales. In this paper we analyze the viability of a set of inflationary potentials in both the context of the collapse proposal and within the standard inflationary framework. For this, we perform a statistical analysis using recent CMB and BAO data to obtain the prediction for the scalar spectral index  $n_s$ in the context of a particular collapse model: the \textit{Wigner} scheme. The predicted $n_s$ and the tensor-to-scalar ratio $r$ in terms of the slow roll parameters is different between the collapse scheme and the standard inflationary model. For each potential considered we compare the prediction of  $n_s$ and $r$ with the limits established by  observational data in both pictures. The result of our analysis shows  in most cases a difference in the inflationary potentials allowed  by the observational limits in both frameworks. In particular,  in the standard approach the more concave a potential is, the more is favored by the data. On the other hand, in the \textit{Wigner} scheme, the data favors equally all type of concave potentials, including those at the border between convex and concave families.

\end{abstract}

\maketitle

\section{Introduction}

According to most recent data reported by the \textit{Planck} mission \cite{planck2015,planck2015inflation,planck2015likelihoods}, the early Universe is consistent with the description provided by cosmic inflation, which assumes an accelerated expansion of the primordial Universe \cite{starobinsky1980,guth,linde,albrecht}. In the simplest scenario, the matter driving the inflationary stage is characterized by a single scalar field, called the inflaton, with canonical kinetic term minimally coupled to gravity \cite{mukhanov1992}.  Moreover, the standard paradigm considers that the inflationary expansion amplifies the quantum fluctuations of the scalar field and converts them into classical perturbations which leave their imprint as temperature and polarization anisotropies in the Cosmic Microwave Background (CMB) \cite{mukhanov81,mukhanov2,starobinsky1982,guth1982,hawking2,bardeen2}. The observational data from  CMB anisotropies constrain the parameters associated to the spectra of primordial fluctuations. Those parameters characterize the amplitude and shape of the scalar/tensor spectra. In particular, \textit{Planck} 2015 data \cite{planck2015inflation} yield the values $\ln(10^{10} As) = 3.094 \pm 0.034 $ and $n_s = 0.9645 \pm 0.0049$ at  68\% confidence level corresponding to the scalar amplitude $A_s$ and spectral index $n_s$ respectively.\footnote{The values are obtained from Table 3 of Ref. \cite{planck2015inflation} considering only temperature (TT), plus polarization (TE,EE) for high multipoles and (TE,EE,BB) for low multipoles.} On the other hand, a measurement of the tensor amplitude $A_t$ and spectral index $n_t$ requires the B--mode polarization of the CMB,  which has not been detected. In fact, it is not known if the tensor spectrum is consistent with a perfect scale invariant spectrum or exhibits some degree of tilt. Regarding the tensor spectrum amplitude, current data can only establish an upper bound, this information is encoded in the so called tensor--to--scalar ratio $r$. The joint analysis of the BICEP2/Keck Array and \textit{Planck} 2015 data set the bound $r<0.12$ at 95\% confidence level \cite{dust5}. 

In order to achieve an inflationary expansion, the potential energy of the inflaton must dominate over its kinetic energy. If there is a region in the potential which is sufficiently flat and the inflaton is located in that region, the accelerated expansion is known as slow roll inflation.  Given observational constraints and theoretical predictions for the inflationary parameters, namely $A_s, n_s$ and $r$, one can determine the specific inflationary potentials consistent with the data (see \cite{baumannbook} for a review). Furthermore, it is also argued that, in order to analyze which potentials are allowed by observations, not only inflation has to be considered but the reheating era as well \cite{jmartinreheating,jmartinreheating2}. A comprehensive list of potentials have been analyzed in such a manner in Refs. \cite{jmartin2013,jmartin2014}  based on the slow roll inflationary model. Those analysis resulted in precise constraints, allowed by the data, on the parameters characterizing each type of potential. The importance of finding out the specific shape of the inflationary potential arises because inflation is supposed to take place at very high energies ($\sim 10^{15}$ GeV) in a regime unreachable by particle accelerators. Hence, the knowledge about the potential and its characteristics can contribute to our understanding of physical processes at such scales. 

In spite of the successful predictions made by the inflationary paradigm, there exists an issue in the standard lore of the model. In particular, there is no consensus on the physical mechanism which transforms quantum fluctuations of the inflaton into actual real inhomogeneities that eventually become imprinted in the CMB anisotropies. This {particular issue} is usually known as the quantum to classical transition of primordial perturbations \cite{grishchuk,grishchuk1992,polarski,lesgourgues1996,grishchuk1997,kiefer}.\footnote{\label{foot2}   { We consider that characterizing the problem as the ``quantum to classical transition" is not completely accurate.  The fundamental description is always quantum mechanical, for instance, quantum mechanics is also valid for macroscopic systems. However, for some physical systems there are certain conditions that allow us to describe specific quantities, to a satisfactory accuracy, by their quantum expectation values. Those can then be identified with their classical counterparts.} }   {Various attempts that have helped to obtain a major understanding of such an issue have been proposed,} the most popular ones are based on decoherence \cite{grishchuk,polarski,lesgourgues1996,grishchuk1997,kiefer,kiefer2006,egusquiza1997,burgess2006}, many--worlds interpretation of quantum mechanics \cite{nomura2011,nomura2011b,mukhanovbook} and evolution of the inflaton's vacuum state into a squeezed state \cite{grishchuk1992,polarski,albrecht1992} or some combination of those   {(in Refs. \cite{shortcomings,LLS} the interested reader can find our posture on the prior proposals). } Nevertheless, a common pragmatic view is to argue that whatever resolves the quantum--to--classical transition of primordial quantum fluctuations, the usual predictions remain unchanged. We think such a view is misguided and in fact, as we will show in the present work, when facing the aforementioned problem and finding a possible solution, the predictions do change (evidently new predictions must be consistent with the data).    {In particular, we will show in Sec. \ref{collapse_review} that the predictions obtained in our model regarding the shape and spectral index of the scalar power spectrum, as well as the amplitude of the tensor power spectrum, are different from the traditional ones.}\footnote{  {It is worthwhile to mention there exist inflationary models that make use of decoherence to explain the quantum to classical transition, and that also change the standard predictions of the inflationary observables} \cite{vennin2018a,vennin2018b}. } 

The orthodox interpretation of quantum mechanics requires a crucial element, namely, an \textit{observer} who performs a measurement using a measurement device. In the early Universe, there is of course no measuring apparatus nor any observer;
consequently, there is nothing that can justifiably be considered as a measurement. One might argue that it is us--humans--on Earth, right now,
who are performing the measurement in question. However, arguing that it
is our observation what leads to the emergence of primordial inhomogeneities,
would be tantamount to saying that we humans create the conditions that
bring about our own existence.\footnote{That is a simply unacceptable closed causal
loop,  as our own existence here and now, requires as a prerequisite, the
formation of galaxies, stars and planets, that must come before even life can
emerge.}  In any case, the issue we have described is directly related to the measurement problem of quantum mechanics.   {In other words, in standard quantum theory, there is no clear definition of what constitutes a measurement (performed by an observer), but this element is required for extracting predictions from the mathematical formalism of quantum mechanics. That, in fact, is achieved through the postulate known as the Born rule.  In the case of the early Universe (and in any cosmological context) the trouble of defining unambiguously such a measurement, i.e. the measurement problem, appears in an exacerbated manner.  }

  {In our view, a more precise characterization of the  problem regarding the origin of primordial inhomogeneities and anisotropies can be summarized as follows.} Starting from a completely homogeneous and isotropic initial setting, characterizing both the vacuum state of the inflaton and the spacetime background, the evolution given by inflationary dynamics, somehow transmutes the initial setting into a final one that is inhomogeneous and anisotropic.  Obviously, this is not simply the result of quantum unitary evolution, since, in this case, the dynamic does not break the initial symmetries of the system. In addition, quantum fluctuations cannot be taken as indicating the existence of inhomogeneities and thus cannot be taken as characterizing them either. In the orthodox interpretation of quantum mechanics, quantum fluctuations are only fluctuations on measurements performed by an observer; the state of the system, and thus its symmetries, are encoded in the state vector or wave function. For the inflaton's vacuum, the symmetries are homogeneity and isotropy.

%

One proposal to deal with the above problem is to invoke an objective (i.e. without observers or measurement devices) collapse of the wave function, corresponding to the inflaton, which can break the  quantum state's initial symmetries \cite{pss,pedrocsl}. The proposal was inspired in early ideas by R. Penrose and L. Di\'osi  \cite{penrose1996,diosi1987,diosi1989} which regarded the collapse of the wave function as an actual physical process (instead of just an artifact of the description of Physics) and it is assumed to be caused by quantum aspects of gravitation. Furthermore, the application of an objective reduction of the wave function to the inflationary Universe has been analyzed  by several authors within different frameworks \cite{jmartin,hinduesS,hinduesT,LB15,magueijo2016,finl}.  The way we treat the collapse process is by assuming that at a certain stage during the inflationary epoch there is an induced jump in a state describing a particular mode of the quantum field, in a manner that is similar to the quantum mechanical reduction of the wave function associated with a measurement, but with the difference that in our scheme no external measuring device or observer is called upon as triggering such collapse.   {The self-induced collapse proposal could be regarded as an alternative explanation to the ``quantum to classical transition" occurring at the time of collapse, which in principle can be any time during the inflationary expansion (however see footnote$^{\textrm{\ref{foot2}}}$).  } The issue that then arises concerns the characteristics
of the post-collapse quantum state. In particular, what determines the expectation values of the field and momentum conjugate variables for the after-collapse state. Previous works by people in our group have extensively
discussed both the conceptual and formal aspects of that problem \cite{pss,shortcomings,LLS,adolfo2008,leon2010a,leon2010,ts,benito,pedro2018}, and the present manuscript will not dwell further into those aspects, except for a very short review. The observational consequences of our proposal have also been analyzed in previous works. Specifically, we have found that the self--induced collapse proposal implies a different prediction with respect to the standard one for the shape and spectral index corresponding to the scalar power spectrum \cite{claudia,pia,micol,micol2}. Additionally, the predicted  amplitude of the tensor power spectrum is very small generically (i.e. undetectable by current experiments), that is, the B--mode polarization spectrum is strongly suppressed \cite{lucila,lmosshort,lmosbig}.\footnote{ The prediction of a  strong suppression of the B--modes amplitude was obtained in the context of semiclassical gravity. If the inflaton's quantization is performed using the Mukhanov-Sasaki variable, as in the standard inflationary model, there is no such suppression.}

Motivated by the fact that  predictions of the inflationary parameters are different between the standard inflationary model and our proposal, in the present work, we have analyzed the feasibility of various inflationary potentials within the self--induced collapse framework. These potentials are of the single slow roll inflation type. Also, we have made use of the public code ASPIC\footnote{The ASPIC name stands for ``Accurate Slow-roll Predictions for Inflationary Cosmology", the library is publicly available at \url{http://cp3.irmp.ucl.ac.be/~ringeval/aspic.html}. } \cite{jmartin2013} which contains 74 slow roll inflationary potentials. However, we considered only 10 potentials of the one parameter kind. The criteria to select the potentials is based on their popularity among cosmologists and the ones that are consistent with the conceptual basis of our model. The output of the ASPIC code allowed us to express our new set of predictions, corresponding to the inflationary parameters, in terms of the potential's parameters. {On the other hand, we performed a statistical analysis using recent CMB \cite{planck2015likelihoods} and Baryon Acoustic Oscillations (BAO) data to obtain the limits on the inflationary parameters in the context of the collapse models.  In such way, we were able to compare  the new predicted inflationary parameters with the observational constraints for those same parameters}.    As a consequence, we can find a range of values for the potential's parameters that are consistent with the data. The novel aspect of this work is that inflationary potentials that were disfavored by observations in the standard framework  become viable within the self--induced collapse framework using the same data.

The paper is organized as follows: In Sec. \ref{collapse_review}, we provide a brief review of the collapse proposal based on \wig $ $ collapse scheme. We also present there the theoretical results that will be of interest for the rest of the paper. In Sec. \ref{potentials}, we present the steps that we will follow regarding our analysis, which involves the observational data and the theoretical predictions in terms of each inflationary potential. In Sec. \ref{results} we present the results of our analysis and the constraints on the parameters of each inflationary potential considered.  We also introduce the computational tools and the data set used in our analysis. Finally, in Sec. \ref{conclusions}, we summarize the main results of the paper and present our conclusions.

\section{Inflation and the \textit{Wigner} collapse scheme}\label{collapse_review}

The objective of this section is to present the results of previous works that will be  of interest for the article's purpose. We only provide a brief review of the collapse inflationary model; thus, there is  no original content in this section. For a complete analysis and motivation we invite the reader to consult our past works (in particular see Refs. \cite{pss,LLS,shortcomings,adolfo2008,pia}).

\subsection{General framework of the self-induced collapse proposal}

Before addressing the self--induced collapse proposal and its connection with inflation, we present our view regarding the relation between gravitational degrees of freedom and matter fields.   {In particular, we will rely on the semiclassical gravity approximation. The crucial feature of this framework is to provide a quantum characterization of the matter fields only, while the metric degrees of freedom remain classical. This approach contrast with the usual procedure in which metric and matter fields perturbations are quantized. In the ensuing paragraphs we expose some arguments which suggest that it is not completely settled that the standard approach is the only option to follow. }

  {There is of course indisputable evidence of the quantum nature of matter, from which it follows that a theory of gravity that acknowledges the quantum character of matter, unlike general relativity,  is necessary. However, even if we  agree that, at the fundamental level, gravitation itself is quantum mechanical in nature, that does not automatically mean that the metric degrees of freedom are the ones that need to be treated quantum mechanically. There are various arguments suggesting that the spacetime geometry might emerge from deeper, non-geometrical and fundamentally quantum mechanical degrees of freedom (see e.g. Refs. \cite{Em1,Em2,Em3,Em4,Em5}); and, just as one does not directly quantize macroscopic, thermodynamic variables, one would not think of quantizing the metric if it were non-fundamental.}

  {Classical gravity is a good effective field theory and one may say that it is straightforward to quantize its linear perturbations. But the fact that a theory is a good effective description classically does not guarantee that quantizing it in a canonical fashion will yield something that accurately describes nature. For instance, nobody believes that quantizing the heat equation is a reasonable thing to do (even though the equation provides a good effective theory). Similarly,  few people would think that quantizing sound waves in the air (in contrast with waves in a solid) is something that would yield reasonable results. The important point is that, in the end, it is always experiments that determine the correct answer. In the particular case of the spacetime metric, the fact is that we simply do not know for sure as we do not have anything as an established theory of quantum gravity.  In  any event, before  such  theory becomes  available   and   before  definite  experimental evidence,  we  simply do not know with absolute certainty if one should canonically quantize the metric perturbations or not.}

  {We agree that small fluctuations around a classical solution can be quantized (in a technical sense) and specially if the theory is represented by a quadratic action. However, it is crucial to notice that there are serious issues that would arise when attempting to give a physical interpretation of the obtained results. For instance, what is the meaning of a spacetime that is in a state of quantum superposition of two states sharply peaked about two spatial metrics? One could claim that such situation is no different than what we face with en electric field. Well, maybe it is, but maybe it is not. In the case of an electric field on a fixed spacetime, we at least know what the superposition implies regarding measurements of the electric field at a certain point. In contrast, with the metric superposition, we encounter the added difficulty of not even knowing how to identify spacetime points. In any case, as we argued above, we cannot know whether or not quantizing the metric is the way to obtain correct description of nature. We simply cannot know that in absence of an established theory of quantum gravity. }

  {In our view, everything ought to  be  described quantum mechanically at its  basic level. However,  since there is not a complete and workable quantum theory of gravity, one can rely on the  semiclassical gravity framework and take it as an approximation, in the appropriate regime, to relate the degrees of freedom of gravity and matter. In fact, the semiclassical framework has provided a suitable approximation when one wants to  take into account quantum effects provided by matter but the gravitational sector can still be characterized by General Relativity, e.g. to derive the thermal radiation emitted by black holes (i.e. Hawking's radiation). On the other hand, we equally expect that,  in the quantum gravity theory, one will be able to find many situations in which the semiclassical Einstein's equations would be completely inappropriate; but it seems quite likely that those would correspond to situations where the concept of spacetime itself becomes meaningless. In the case presented in the paper, we are using the semiclassical gravity approximation as suitable description in which one can observe (as will be shown in subsection 2.4) how the curvature perturbation (which again is always  classical) is born from the quantum collapse.} \footnote{We note that although the prevailing perception is that combining a classical theory of gravity with a quantum theory of matter would be inconsistent,  the issue  is  still  an open one.  In fact, there are a number of arguments in the literature against the viability of a half-and-half theory. Nevertheless, as  shown  in Refs. \cite{unviable-semicalsssical1,unviable-semicalsssical2,unviable-semicalsssical3,unviable-semicalsssical4,unviable-semicalsssical5,unviable-semicalsssical6,unviable-semicalsssical7}.  none of those arguments are really conclusive.}

Semiclassical gravity is encoded in Einstein's semiclassical equations
\begin{equation}
R_{ab} - (1/2) R  g_{ab} = 8 \pi G \bra \hat T_{ab} \ket.
\end{equation}

Having adopted a point of view whereby the spontaneous collapse is the underlying mechanism behind the breakdown of homogeneity and isotropy in the inflationary epoch, and given that spontaneous collapse is, at the basic level, a modification of the temporal quantum evolution, it seems natural to work in a setting where time and, more generally speaking, spacetime notions, appear as playing their standard roles. That is, it seems natural to work in contexts where the spacetime metric is taken as well--defined and subject to a classical treatment. Thus, our program seems best framed where the spacetime metric is treated classically while matter fields are described quantum mechanically. 

We of course do not claim that this is the only choice. As a matter of fact, one might choose instead to rely on the standard idea of quantizing perturbations of both matter and metric fields, and to incorporate in that setting the collapse proposal.\footnote{Such approach is followed, for instance, in Refs.\cite{gabriel,jmartin,hinduesS,hinduesT,LB15,mauro,magueijo2016}.} We note, however, that in the absence of a workable theory of quantum gravity (and considering the fact that canonical approaches to quantum gravity invariably face the ``problem of time"), such approach can at best be attempted in a perturbative setting (where the causal structure is taken as that of the unperturbed background spacetime metric). On the other hand, using a spontaneous collapse within the semiclassical setting, is, in principle, susceptible to a non--perturbative treatment using, for instance, the formalism developed in Refs.\cite{ts,pedro2018}.

Let us now focus on the theory. The action of a single scalar field minimally coupled to gravity is
\begin{eqnarray}\label{action0}
S[\phi,g_{ab}] &=& \int d^4x \sqrt{-g} \bigg[ \frac{1}{16 \pi G} R[g] \nn
&-& \frac{1}{2}\nabla_a \phi \nabla_b \phi g^{ab} - V[\phi] \bigg]. 
\label{actioncol}
\end{eqnarray}
Next one splits the scalar field and metric  into background plus perturbations, i.e. $g_{ab} = g_{ab}^{(0)} + \delta g_{ab}$ and $\phi (\x,\eta) = \phi_0 (\eta) + \dphi (\x,\eta)$.

The background metric is described by a spatially flat FRW spacetime. In conformal coordinates, the components of the background metric are $ g_{\mu \nu}^{(0)} = a(\eta) \eta_{\mu \nu}$, with $a(\eta)$ the scale factor,  $\eta $  the conformal cosmological time  and  $\eta_{\mu \nu}$ the components of the Minkowskian metric. The Hubble slow roll parameters, are defined as $\epsilon_1 \equiv 1- {\mH'}/{\mH^2}, \epsilon_2 \equiv {\epsilon_1'}/{\mH \epsilon_1}$, here $\mH \equiv aH$ is the conformal Hubble factor and primes over functions $f'$ denote derivative with respect to $\eta  $. Using Friedman's equation for the background, one can write $\mH$ as a function of $\phi_0$. In that case, if the slow roll parameters are small, i.e. $\epsilon_1 \ll 1$ and $\epsilon_2 \ll 1$, then  slow roll inflation is guaranteed. The slow roll approximation implies $3 \mH \phi_0'  \simeq -a^2 \partial_\phi V$ and $ \mH^2 \simeq (8 \pi G/3) a^2 V $.

Moreover,  the slow roll parameters can be related to the inflaton's potential through,
\begin{subequations}\label{PSR}
	\begin{equation}
	\epsilon_1 \simeq \frac{M_P^2}{2} \left( \frac{\partial_\phi V}{V}\right)^2, 
	\end{equation}
	\begin{equation}
	\epsilon_2 \simeq 2 M_P^2  \left[  \left( \frac{\partial_\phi V}{V}\right)^2 -  \left( \frac{\partial_{\phi \phi} V}{V}\right) \right].
	\end{equation} 
\end{subequations}
where $M_P^2 \equiv 1/(8\pi G)$ is the reduced Planck's mass.

  {At this point we should focus on a particular subject of cosmological perturbation theory, i.e. the gauge issue.  In classical GR perturbation theory, the problem of gauge is connected to the fact  that one  is  comparing two pseudo-Riemannian manifolds:  the  background $(M, g_{ab}, \phi)$ (that is manifold, metric and  field)  and a perturbed one $(M', g'_{ab}, \phi')$. Moreover,  when focusing on the perturbations: $ \delta g_{ab} =  g'_{ab} - g_{ab}$  or $ \delta \phi = \phi' - \phi$, one  must deal with the  arbitrariness of  choice of which point of $M'$ is to be identified with  each one of $M$ (in order to  evaluate the differences  defining the aforementioned $\delta$-fields ). There are  two  approaches  used   generically to deal  with this problem:}

\begin{enumerate}
	\item   {To fix the gauge, }
	
	\item   {To  work  with  the so  called  ``gauge invariant quantities".}
\end{enumerate}

  {Approach 2  is  traditionally favored in this  particular branch of cosmology, and is  based  on the fact that  certain   combinations of metric components (associated  with a previous selection of  preferential coordinates  correlated  with the  homogeneous and isotropic  background) and of  the field variations,  have the property of being invariant under infinitesimal  coordinate  transformations.  We stress this fact  because confusion might arise  in these matters, and mistakenly lead one to entertain the notion that somehow geometry and  matter fields become ``inexorably mixed"  to the degree  that  they only acquire meaning  in those, so called,  gauge invariant combinations. This is not correct,  as  geometry and matter fields  are  quite different physical objects  (one  can be  measured  for instance  by  analyzing the  geodesic deviations  in a neighborhood,  and the other  by placing  a suitable   detector that interacts  with the  fields in question).}

  {In any  event, as we  are treating  geometry and fields in a very different  fashion in our approach (the perturbations of  the inflaton field is treated quantum mechanically--QFT in curved spacetime--, while the  metric  description of geometry--which we are  regarding as emergent--is  treated  classically), we cannot hope to use the second approach.  That is we choose Approach 1 and work in a fixed  gauge. Nonetheless, we  expect  gauge invariance of our  answers  to the  same  degree that  one  has  gauge invariance of any analysis  carried  out in a particular gauge, and  we  do expect all  those  choices  that represent real  physical  alternatives, to modify the results  accordingly. In particular,  a notion of the  ``time" at  which collapse  of certain mode takes place,  might change with a different of gauge (this ``time" of course being  just a index  or label, with  no physical  significance). However, physically relevant quantities, e.g.  the actual mean density, as  hypothetically  measured by comoving observers at the  onset of the collapse,  would not depend on the  gauge. We will get back to this issue  after a  small  but necessary digression. }

  {The choice of  gauge  implies that the time  coordinate is attached  to  some specific  slicing of  the  perturbed spacetime, and thus our identification of  the corresponding  hypersurfaces (those of constant time)  as the  ones  associated with the  occurrence of collapses--something deemed as an actual physical  change--turns  what is normally  a simple  choice of  gauge  into a choice of   the distinguished  hypersurfaces,  tied to the putative physical process behind  the collapse. This  naturally leads to  tensions  with the expected   general covariance of a fundamental theory, a  problem that afflicts  all known  collapse  models, and  which in the non-gravitational   settings becomes the issue   of compatibility with  Lorentz  or Poincare invariance of the proposals.  We must acknowledge that this generic problem  of  collapse  models is indeed  an  open issue for the present approach. One  would expect that its  resolution  would be tied to the  uncovering of the  actual  physics  behind  what we treat here as  the  collapse of  the  wave function (which we  view as a merely an effective description). As has been argued in related  works, and  in following ideas originally  exposed  by Penrose \cite{penrose1996},   we hold that the physics  that  lies behind  all  this,  ties the  quantum  treatment of gravitation  with the foundational issues  afflicting quantum theory in general, and in particular those  with connection  to the ``measurement problem.'' }

  {Let us now turn back to the connection of those issues with the choice of gauge in our approach:  The ``choice of  gauge" determines  among  other things which  hypersurfaces of the  perturbed spacetime  are  labeled  as  surfaces of constant time (where ``time" would be that preferred  time-like coordinate of the  unperturbed FRW  spacetime, defined  trough  the pull--back  on the perturbed  spacetime), and  as  such, in our situation, they determine  on which hypersurfaces  does the collapse occur (we have  been working under the assumption that collapse occurs  on these ``equal time hypersurfaces").   Thus, this choice is no longer just a  gauge  choice,  but an actual  assumption about which hypersurfaces  are  the ones  one  can associate  with the quantum collapse or reduction. This  seems inescapable, and  in fact a desired feature.  After all, if the physical condition of a system  is to  be represented by a quantum state, and if the initial  state is homogeneous and isotropic, and  the  final one is not. Then, any development  interpolating between the two, will require the selection  of a time---in our case  hypersurface---where the  transition occurs (perhaps a  more developed  method  will involve  a full  series  of times, or  a continuous period  of time).  In this  sense,  once  we identify  the surface of collapse,  a  change of gauge  would  involve a  complicated  re--specification of the ``times"  at which each  comoving  observer crosses the collapse hypersurface. These arguments have been formally developed in what is named the Semiclassical Self--consistent configuration (SSC) framework, see Refs. \cite{ts,pedro2018} for a full analysis.  In the present article, we make use of the results of the aforementioned works, without dwelling to much into the full mathematical formalism.}


  {Given the discussion above, we choose to work in the longitudinal (or \textit{Newtonian}) gauge. The advantage of working with this gauge is that the action at second order involving the matter and metric perturbations is mathematically equivalent as the one using gauge invariant quantities. Therefore, we can be certain that the field perturbations are actual physical degrees of freedom and not pure gauge.} Thus, assuming no anisotropic stress and working in the longitudinal gauge, the line element associated to scalar metric perturbations is
\begin{equation}
ds^2 = a^2(\eta) \left[ - (1+2\Psi) d\eta^2 + (1-2 \Psi) \delta_{ij} dx^i dx^j \right].
\end{equation}

As we have mentioned, the semiclassical approximation implies that only the inflaton will be described by a quantum field theory; in contrast, the metric (background and perturbations) is always  classical.   {We will focus first on the classical dynamics of the perturbations and then proceed to the quantum theory of the inflaton's inhomogeneous part  $\dphi (\x,\eta)$}

  {In  Appendix A of Ref. \cite{pia} it is shown that combining the perturbed Einstein  equations with components $\delta G^0_0 = 8 \pi G \delta T^0_0$, $\delta G^0_i = 8 \pi G \delta T^0_i$, $\delta G^i_j = 8 \pi G \delta T^i_j$  and the slow-roll motion  equation, one obtains:
\begin{equation}\label{master00}
\nabla^2 \Psi +\mu \Psi =  4 \pi G \phi_0' \dphi'
\end{equation}
where $\mu \equiv \mH^2-\mH'$. In Fourier space, and applying the slow roll equations once again, Eq. \eqref{master00} reads 
\begin{equation}\label{master1}
\Psi_{\nk} (\eta) = \sqrt{\frac{\epsilon_1}{2}} \frac{\mH}{M_P (k^2-\mu)}  \dphi_{\nk}' (\eta)
\end{equation}}

  {Generalizing the above equation using the semiclassical gravity approximation we have}
\begin{equation}\label{master1}
\Psi_{\nk} (\eta) = \sqrt{\frac{\epsilon_1}{2}} \frac{\mH}{M_P (k^2-\mu)} \bra \hat  \dphi_{\nk}' (\eta) \ket.
\end{equation}

  {It is important to mention that we are not indicating that there are inhomogeneities of any definite size in the inflationary Universe, but merely we are analyzing the dynamics if such inhomogeneities existed. In particular, the fundamental description corresponding to the inhomogeneous part of the matter field $\dphi$ is dealt at the quantum level, where $\dphi$ is a quantum field in a given quantum state. In fact that will be our next task. But before analyzing the QFT of $\dphi$, we would like to mention that Eq. \eqref{master1} was obtained by working in the longitudinal gauge. However, as has been shown in Ref. \cite{pia} the same equation is obtained by working with gauge invariant quantities. Therefore, we can assure that Eq. \eqref{master1} truly reflects the connection between the physical degrees of freedom associated to matter and geometry. In particular, in the longitudinal gauge,  $\Psi$ is the curvature perturbation, i.e.  it is the intrinsic spatial curvature on hypersurfaces on constant conformal time for a flat Universe. }

\subsection{Quantum theory of perturbations and  \textit{Wigner's} collapse scheme}

  {Next, we focus on the quantum description of $\dphi$.} Our  treatment is based on the quantum theory of  $\dphi (\x,\eta)$ in a curved background described by a  quasi--de Sitter spacetime. It is convenient to  work with the  rescaled field variable $y=a\dphi$.    {Expanding the action \eqref{action0} up to second order in the $y$ variable, we obtain $\delta^{(2)} S = \int d^4x \delta^{(2)} \mathcal{L}$, where}
\begin{eqnarray}\label{action2}
\delta^{(2)} \mathcal{L} &=& \frac{1}{2} \bigg[ y'^2   - (\nabla y)^2 + \left( \frac{a'}{a} \right)^2 y^2      \nn
&-&2 \left( \frac{a'}{a} \right) y y' - y^2 a^2 \partial_{\phi \phi} V \bigg].
\end{eqnarray}
  {Note that in $\delta^{(2)} S$ there are no terms containing metric perturbations since it is only after the self--induced collapse that the spacetime is no longer homogeneous and isotropic. }

Next, the field $y$ and the  canonical conjugated momentum   $\pi \equiv \partial \delta^{(2)}  \mathcal{L}/\partial y' = y'-(a'/a)y=a\dphi'$ are promoted to quantum operators so that they satisfy the following equal time commutator relations:  $[\hat{y}(\x,\eta), \hat{\pi}(\x',\eta)] = i\delta (\x-\x')$ and 
$[\hat{y}(\x,\eta), \hat{y}(\x',\eta)] = [\hat{\pi}(\x,\eta), 
\hat{\pi}(\x',\eta)] = 0$. We can expand  the field operator in discrete Fourier's modes (at the end of the calculation we take the limit $L\to \infty$ and $\nk$ continuous) 
\begin{equation}
\hat{y}(\eta,\x) = \frac{1}{L^3} \sum_{\nk} \hat{y}_{\nk} (\eta) e^{i \nk \cdot \x},  
\end{equation}
with an analogous expression for $\hat{\pi}(\eta,\x)$. Note that  the sum is over the wave vectors $\vec k$ satisfying $k_i L=2\pi n_i$  for $i=1,2,3$ with $n_i$ integer and $\hat y_{\nk} (\eta) \equiv y_k(\eta)  \ann_{\nk} + y_k^*(\eta) \cre_{-\nk}$ and  $\hat \pi_{\nk} (\eta) \equiv  g_k(\eta) \ann_{\nk} + g_{k}^*(\eta) \cre_{-\nk}$, with $g_k(\eta) = y_k'(\eta)  - \mH y_k (\eta)$.    {The field's mode equation of motion is }
\begin{equation}\label{ykmov2}
y''_k(\eta) + \left(k^2 - \frac{a''}{a} +  a^2 \partial_{\phi \phi} V \right)  y_k(\eta)=0.
\end{equation}
  {At first order in the slow roll parameters, we have}
\begin{equation}
 -\frac{a''}{a} +  a^2 \partial_{\phi \phi} V\simeq  \frac{-2+3 \epsilon_1 - (3/2)\epsilon_2}{\eta^2}.
\end{equation}

The selection of $y_k(\eta)$ reflects the choice of a vacuum state for the field. We proceed as in standard fashion and choose the  Bunch-Davies vacuum: 
\begin{equation}\label{nucolapso}
y_k (\eta) = \left( \frac{-\pi \eta}{4} \right)^{1/2} e^{i[\nu + 1/2] (\pi/2)}
H^{(1)}_\nu (-k\eta), 
\end{equation}
where $ \nu \equiv 3/2 -\epsilon_1 + \epsilon_2/2$  and $H^{(1)}_\nu (-k\eta)$ is the Hankel function of first kind and order $\nu$. 


  {To describe the collapse of the state associated to the scalar field, we use the decomposition of the field into modes. It is necessary that these modes are independent, i.e. that they give a corresponding decomposition of the field operator into a sum of commuting ``mode operators,'' an orthogonal decomposition of the one-particle Hilbert space and a direct-product decomposition for the Fock space. Furthermore, we require that the initial state of the field is not an entangled state with respect to this decomposition, i.e. it can be written as a direct product of states for the mode operators. This ensures that the notion of ``collapse of an individual mode'' will make sense. }

  {Let us be more precise:} the collapse hypothesis assumes that  at a certain time $\tc$ the part of the state characterizing the mode $k$ randomly ``jumps'' to a new state, which is no longer homogeneous and isotropic. The collapse is considered to operate similar to an imprecise ``measurement'', even though there is no external  observer or detector involved. Therefore, it is reasonable to consider Hermitian operators, which are susceptible of a direct measurement in ordinary quantum mechanics. Hence, we separate $\hat y_{\nk} (\eta)$ and $\hat \pi_{\nk} (\eta)$ into their real and imaginary parts $\hat y_{\nk} (\eta)=\hat y_{\nk}{}^R (\eta) +i \hat y_{\nk}{}^I (\eta)$ and $\hat \pi_{\nk} (\eta) =\hat \pi_{\nk}{}^R (\eta) +i \hat \pi_{\nk}{}^I (\eta)$, such that the operators $\hat y_{\nk}^{R, I} (\eta)$ and $\hat \pi_{\nk}^{R, I} (\eta)$ are  Hermitian operators. Thus, 
\begin{subequations}\label{operadoresRI}
	\begin{equation}
	\hat{y}_{\nk}^{R,I} (\eta) = \sqrt{2} \textrm{Re}[y_k(\eta) \hat{a}_{\nk}^{R,I}],
	\end{equation}
	\begin{equation}
	\hat{\pi}_{\nk}^{R,I} (\eta) = \sqrt{2} 
	\textrm{Re}[g_k(\eta) \hat{a}_{\nk}^{R,I}],
	\end{equation}
\end{subequations} 
where $\hat{a}_{\nk}^R \equiv (\hat{a}_{\nk} + \hat{a}_{-\nk})/\sqrt{2}$, 
$\hat{a}_{\nk}^I \equiv -i (\hat{a}_{\nk} - \hat{a}_{-\nk})/\sqrt{2}$. 

The commutation relations for the $\hat{a}_{\nk}^{R,I}$ are non-standard $[\hat{a}_{\nk}^{R,I},\hat{a}_{\nk'}^{R,I \dag}] = L^3 (\delta_{\nk,\nk'} \pm 
\delta_{\nk,-\nk'}),$ the $+$ and the $-$ sign corresponds to the commutator with the $R$ and $I$ labels respectively; all other commutators vanish.


  {Now we specify the rules according to which the collapse happens. Again, our criteria is simplicity and naturalness. In particular, we will proceed in a purely phenomenological manner by introducing a general parameterization fo the quantum state after collapse; we will refer to this approach as a \textit{collapse scheme.}  Specifically, the collapse scheme serves to characterize the post--collapse state by the quantum expectation values of the field and its conjugated momentum at the time of collapse. As a consequence of the collapse, those expectation values change from being zero, when evaluated in the vacuum state, to having non--vanishing value in the post--collapse state. The particular collapse scheme used leads to an expression for the post--collapse expectation values, leaving an imprint in the primordial power spectrum (the details will be shown in subsection 2.5)     }

  {We now introduce the \textit{Wigner's collapse scheme} (this scheme was first introduced in \cite{adolfo2008} and subsequently analyzed in great detail \cite{claudia,leon2010,pia,micol,micol2}): In non--relativistic QM, Heissenberg's uncertainty principle indicates that quantum uncertainties of position and momentum operators are  not independent. Specifically, momentum and position of a quantum system cannot  be determined simultaneously and independently. The self-induced collapse acts as a sort of  spontaneous  ``measurement" (of course without relying on observers/measurements devices) of some variable  involving both position and momentum. Consequently,  as suggested  by the uncertainty principle,  the collapse might  involve   correlations between position and momentum.  Generalizing this fact to our inflationary model seems to indicate  that the self-induced collapse could correlate the field $\hat y$ and its conjugated momentum $\hat \pi$.}

  {One possible manner to characterize the correlation is to use Wigner's distribution function. In non-relativistic QM, Wigner's function can be considered, under certain special circumstances, as a probability distribution function for a quantum system, i.e. it allow us to visualize the momentum-position correlations and quantum interferences in ``phase space." For the vacuum state of each mode of the inflaton, the corresponding Wigner's function is a bi-dimensional Gaussian. As a consequence, in this scheme  we will characterize the post-collapse expectation values as:  }
\begin{subequations}\label{esquemawig}
	\begin{equation}
	\bra \hat{y}^{R,I}_{\nk}(\tc)\ket  = x_{\nk}^{R,I} \Lambda_k (\tc) 
	\cos \Theta_k (\tc), 
	\end{equation}
	\begin{equation}
	\bra \hat{\pi}^{R,I}_{\nk}(\tc) \ket = x_{\nk}^{R,I} k \Lambda_k (\tc) 
	\sin \Theta_k (\tc),
	\end{equation}
\end{subequations}
where $x_{\nk}^{R,I}$ is a random variable with a Gaussian  probability distribution function,  centered at zero with  spread one.   { The variable $\tc$ represents the conformal time of collapse.} The quantity $\Lambda_k (\tc)$ represents the major semi--axis of the ellipse in the $\hat{y}$--$\hat{\pi}$ plane where the Wigner function has magnitude 1/2 of its maximum value. The other variable $\Theta_k (\tc)$ is the angle between $\Lambda_k (\tc)$ and the $\hat{y}_{\nk}^{R,I}$ axis. The explicit expressions for $\Lambda_k$  and $\Theta_k $ are very cumbersome and do not contain much information for our present interest.\footnote{The interested reader can consult Refs. \cite{pia,micol2} for the exact expressions of $\Lambda_k (\tc)$, $\Theta_k (\tc)$  and their derivation.    }


\subsection{Emergence of curvature perturbation within the collapse scheme}

  {Here we illustrate how the collapse process generates the seeds of cosmic structure. We proceed by recalling that the conjugated momentum is $\hat \pi_{\nk} = a \hat \dphi'_{\nk} $, therefore Eq. \eqref{master1} can be expressed as}
\begin{equation}\label{master2}
\Psi_{\nk} (\eta) = \sqrt{\frac{\epsilon_1}{2}} \frac{H}{M_P (k^2-\mu)} \bra \hat  \pi_{\nk} (\eta) \ket.
\end{equation}
  {Given that we are working in the longitudinal gauge, then the scalar $\Psi_{\nk}$ represents the curvature perturbation. Moreover, the primordial curvature perturbation $\Psi_{\nk}$ is related to the quantum expectation value   of the conjugated momentum $\bra \hat \pi_{\nk} \ket$. It follows from the  above equation that in the vacuum state $ \bra \hat{\pi}_{\nk}  \ket_{ 0} =0$, which implies $\Psi_{\nk}  =0$, i.e., there are no perturbations of the symmetric  background spacetime. It is only after the collapse has taken place  ($|\Theta\ket \neq |0\ket$)  that $\bra \hat{\pi}_{\nk} \ket_\Theta \neq  0$ generically and $\Psi_{\nk} \neq 0$; thus, the primordial inhomogeneities  and anisotropies arise from the quantum collapse. 
}

  {As we observe from \eqref{master2}, the time evolution of $\Psi_{\nk} (\eta) $ is driven by the dynamics of $\bra \hat \pi_{\nk} (\eta) \ket$ evaluated in the post-collapse state. The corresponding expression  of $\bra \hat \pi_{\nk} (\eta) \ket$ is of the form   }
\begin{equation}\label{masterpi}
\bra \hat \pi_{\nk} (\eta) \ket = F(k\eta,z_k) \bra \hat y_{\nk} (\tc) \ket + G(k\eta,z_k)  \bra \hat \pi_{\nk} (\tc) \ket.
\end{equation}
  {The parameter $z_k$ is defined as $z_k \equiv k \tc$. The deduction of such equation and its explicit function is shown in Appendix B of \cite{pia}. The important point is that $\bra \hat \pi_{\nk} (\eta) \ket$ depends linearly on the expectation values  $\bra \hat y_{\nk} (\tc) \ket$ and $\bra \hat \pi_{\nk} (\tc) \ket$ evaluated at the time of collapse. Those expectations values are the ones characterized by the Wigner's collapse scheme presented in Eqs. \eqref{esquemawig}. }

  {Substituting \eqref{masterpi} in \eqref{master2}, and making use of the Wigner's scheme,  we find the expression for the primordial curvature perturbation (given in the longitudinal gauge)}
\begin{eqnarray}\label{masterpsi}
\Psi_{\nk} (\eta) &=&  \sqrt{\frac{\epsilon_1}{2}} \frac{H}{M_P (k^2 - \mu) } \nonumber 
\\
&\times& X_{\nk} \Lambda_k(z_k) \bigg[ F(k\eta,z_k) \cos \Theta_k (z_k)   \nonumber \\ 
&+& G(k\eta,z_k) k \sin \Theta_k (z_k) \bigg],
\end{eqnarray}
  {where $X_{\nk}\equiv x^R_{\nk} + i x^I_{\nk}$}.

  {Up to this point we have proceeded in with no approximations in order to obtain $\Psi_{\nk} (\eta)$. In the present work, we will be only interested in the case where the time of collapse occurs well before the ``horizon crossing,'' i.e. when the time of collapse satisfies $-k \tc \gg 1 $ or equivalently, with the definition of $ z_k$, when $|z_k| \gg 1$. We remind the reader that before the time of collapse there are no perturbations whatsoever, that is $\Psi_{\nk} = 0$. After the time of collapse, when the primordial perturbation is born, $\Psi_{\nk} (\eta)$ evolves. Hence, for any given mode, $\Psi_{\nk} (\eta)$ originates well inside the horizon, then it continues to evolve until horizon crossing, and finally enters into the super--horizon regime. }

  {In particular, the time evolution of $\Psi_{\nk} (\eta)$ is dictated by the functions $F$ and $G$. The time dependence of those functions are given by linear combination of Bessel's functions $J_{\nu} (k\eta)$ and $Y_{\nu} (k\eta)$ (see Appendix B of Ref. \cite{pia}).  Therefore, when the modes are sub-horizon ($k\eta \gg 1$ ), the curvature perturbation $\Psi_{\nk} (\eta)$ oscillates. For super-horizon modes  ($k\eta \ll 1$) the curvature perturbation is approximated by} 
\begin{equation}\label{psifuerahorizonte}
	\Psi_{\nk} (\eta) \simeq \mA W(z_k)  \eta^{3/2-\nu} k^{-\nu} X_{\nk}, 
\end{equation}	
  {with $\mA$ some amplitude which includes numerical factors, $H$ and $\epsilon_1$; the function $W(z_k)$ is a cumbersome function of the time of collapse $z_k \equiv k\tc$; also  we recall from \eqref{nucolapso} that  $ \nu \equiv 3/2 -\epsilon_1 + \epsilon_2/2$. Furthermore, as shown in Ref. \cite{pia}, the quantity	$\Psi_{\nk} (\eta) $ given by \eqref{psifuerahorizonte} is constant up to second order in slow roll parameters (one needs to take into account that the time dependence appears not only in the factor $\eta^{3/2-\nu}$ but also implicitly in the $H$ and $\epsilon_1$ parameters which are not strictly constant, i.e. they do not describe an exact de--Sitter spacetime). }

\subsection{Primordial scalar power spectrum}

  {Having discussed the origin of the primordial curvature perturbation, we now focus on the scalar power spectrum. The scalar power spectrum in Fourier space is defined as }
\begin{equation}\label{defPS}
\overline{\Psi_{\nk} \Psi_{\nk'}^\star} \equiv \frac{2 \pi^2}{k^3} \mP_s \delta(\nk-\nk'),
\end{equation}
  {where $\mP_s(k)$ is the dimensionless power spectrum. The bar appearing in \eqref{defPS} denotes an ensemble average over possible realizations of the stochastic field $\Psi_{\nk}$. In our approach, the realization of a particular $\Psi_{\nk}$ is given by the self-induced collapse according to the Wigner's scheme. }

  {Using expression \eqref{psifuerahorizonte}, one can compute  $\overline{\Psi_{\nk} \Psi_{\nk'}^\star}$. Furthermore, since  $ \Psi_{\nk} $ is a constant (up to second order in the slow roll parameters), we can evaluate  $ \Psi_{\nk} (\eta)  $ at the time $\eta_* = 1/k_*$ i.e. at the conformal time of horizon crossing corresponding to a particular $k_*$ called the pivot scale. Note that were are following the same method as the traditional one when evaluating the power spectrum at the horizon crossing even if the expression for the curvature perturbation considered was obtained in the super-horizon regime (the error induced is of higher order in the slow roll parameters, see \cite{kinney} for a useful discussion on this subject). Also we further assume that the random variables are uncorrelated, that is, they satisfy, $\overline{x^{R,I}_{\nk} x^{R,I}_{\nk'}} = \delta_{\nk,\nk'} \pm \delta_{\nk,-\nk'} $; the $+$ corresponds  to the real part $x^{R}_{\nk}$ and the $-$ corresponds  to the imaginary part $x^{I}_{\nk}$. Consequently,  $\overline{X_{\nk} X^\star_{\nk'}} = 2 \delta(\nk-\nk')$ (in the continuous limit of $\nk$). }

  {Taking into account the above discussion, it is straightforward to obtain $\overline{\Psi_{\nk} (\eta_*) \Psi_{\nk'}^\star (\eta_*)}$ from \eqref{masterpsi}. That is,    }
\begin{equation}
\overline{\Psi_{\nk} (\eta_*) \Psi_{\nk'}^\star (\eta_*)} = \mA^2 W(z_k)^2  k_*^{-3+2\nu} k^{-2\nu} \delta(\nk-\nk').
\end{equation}
  {From the latter expression and the definition \eqref{defPS}, we can extract the scalar power spectrum}
\begin{equation}
\mP_s(k) = \frac{\mA^2}{2 \pi^2} W(z_k)^2  \left( \frac{k}{k_*} \right)^{3-2\nu}.
\end{equation}

  {Finally, we re-express the obtained power spectrum in a more familiar manner, i.e. (for more technical details see \cite{pia}) }
\begin{equation}\label{Pk}
\mP_{s} (k) = A_s \left( \frac{k}{k_*} \right)^{n_s-1} Q(z_k)
\end{equation}
with $A_s = {H^2}/{8 \pi^2 \epsilon_1 M_P^2}$ (the parameters $H$ and $\epsilon_1$ are evaluated at the horizon crossing of the pivot scale, i.e. at $\eta_* = 1/k_*$) and
\begin{equation}\label{nscol}
n_s-1 = + 2 \epsilon_1 - \epsilon_2
\end{equation}
  {Expression \eqref{Pk} is our predicted scalar primordial power spectrum (at first order in the slow roll parameters), using the \textit{Wigner} collapse scheme (within the semiclassical gravity framework). }

Note that the obtained amplitude $A_s$ is exactly the same as in the standard treatment while the spectral index is different. The traditional prediction of the scalar spectral index is $n_s^{\std}-1 = -2 \epsilon_1 - \epsilon_2$.

Another main difference introduced by the collapse model, as compared with the traditional prediction, is an extra $k$ dependence in the spectrum  reflected in the function
\begin{eqnarray}\label{pwignerdentro}
& & Q(z_k) \equiv \nn
& & \bigg\{  \bigg[ \frac{2\nu}{\zk^{3/2}}  \left( \cos \beta(\nu,|z_k|) - \frac{\sin \beta(\nu, 
	|z_k|) }{2|z_k|} \frac{\Gamma(\nu+3/2)}{\Gamma(\nu - 1/2)}     \right)  \nn
&-& \left( \sin \beta(\nu,|z_k|) + \frac{\cos \beta(\nu, |z_k|) }{2|z_k|} 
\frac{\Gamma(\nu+5/2)}{\Gamma(\nu + 1/2)}     \right)            \bigg] \cos 
\Theta_k   \nonumber \\
&+&  \left[ \cos \beta(\nu,|z_k|) - \frac{\sin \beta(\nu, |z_k|) }{2|z_k|} 
\frac{\Gamma(\nu+3/2)}{\Gamma(\nu - 1/2)}     \right] \sin \Theta_k  \bigg\}^2, \nn
\end{eqnarray} 
where $\Gamma(x)$ is the Gamma function, $\nu=2 - {n_s}/{2}$, $\beta(\nu,|z_k|) \equiv |z_k| - (\pi/2) (\nu+1/2)$ and $\tan 2\Theta_k  \simeq -4/3\zk$.

The model's parameter is the time of collapse $\tc$ or equivalently the quantity $z_k \equiv k \tc$. We parameterize the time of collapse as
\begin{equation}
\tc = \frac{A}{k} + B.
\label{etac}
\end{equation}
The motivation regarding that specific parameterization has been discussed in previous works \cite{pss,adolfo2008,claudia,pia,micol2}. We can see that  if $B=0$ then $Q(z_k)$ is a constant, and our model's power spectrum has the same shape as the standard prediction, i.e. $\propto k^{n_s-1}$. 

Another important aspect is that, in order to obtain the spectrum \eqref{Pk}, the approximation $|z_k| \gg 1$ was used. This means,  we are assuming that the time of collapse takes place before the so called ``horizon crossing,'' at least for modes that contribute the most to the observed CMB anisotropies. In other words, we take $A$ and $B$ such that $-k \tc \gg 1$ with $k$ between $10^{-6}$ Mpc$^{-1}$  and $10^{-1}$ Mpc$^{-1}$.

\subsection{Primordial tensor power spectrum}

  {Regarding tensor perturbations of the metric, the line element corresponding to the perturbed flat FRW metric (at first-order) is given by}
\begin{equation}
ds^2 = a(\eta)^2 [-d\eta^2 + (\delta_{ij}+ h_{ij} ) dx^i dx^j   ].
\end{equation}
  {Therefore, Einstein's perturbed equations (at first-order)  with $i,j$ components yield}
\begin{equation}\label{tensormodes}
h_{ij}'' + 2\mH h_{ij}' - \nabla^2 h_{ij} = 0
\end{equation}
  {In the traditional inflationary scenario, the metric perturbations are quantized. Hence, whatever drives the quantum to classical transition in the scalar sector, must also do the same for tensor modes. As a consequence, one can associate a tensor power spectrum to the quantum two-point function (in Fourier's space) $\bra 0 | \hat h^i_j (\nk,\eta) \hat h^j_i (\nk',\eta) |0\ket  $  \cite{mukhanovbook}.  In fact, in the standard approach,  one expects that the tensor power spectrum acquires an amplitude that in principle can be detected in the CMB B--mode polarization spectrum. }

  {On the other hand, in our proposal based on the semiclassical framework,  Eq. \eqref{tensormodes} does not contain matter sources; this contrast with the scalar perturbation in which $\Psi_{\nk}$ is sourced by the quantum expectation value $\bra \hat \pi_{\nk} \ket $, see Eq. \eqref{master2}. Therefore, in our approach, there are no primordial tensor modes at first order in the perturbations. Thus, in the self--induced collapse proposal based on the semiclassical gravity approximation, we need to consider second-order perturbation theory in order to deal with the primordial gravitational waves.}

  { The analogous expression to \eqref{master2} obtained from Einstein's perturbed equations (at second-order) is (see Refs. \cite{lucila,lmosbig})   }
\begin{equation}\label{tensormodescolapso}
h_{ij}'' + 2\mH h_{ij}' - \nabla^2 h_{ij} = 16 \pi G \bra ( \partial_i \hat \dphi ) \ket \bra ( \partial_j \hat \dphi ) \ket^\textrm{tr-tr}, 
\end{equation}
  {where the superscript tr-tr stands for the transverse-traceless part of the expression. Note that as before, even though $\bra ( \partial_i \hat \dphi ) \ket \bra ( \partial_j \hat \dphi ) \ket$ vanishes when evaluated in the Bunch-Davies vacuum, it will become non-vanishing in the quantum state of the field that results from the spontaneous collapse.}

  {The fact that the scalar metric perturbations are seeded by the linear terms in perturbations of the scalar field [see. Eq. \eqref{master2}], while the tensor perturbations are seeded by quadratic terms, represents a major difference between our approach and the standard one. In particular, an equivalent tensor power spectrum can be obtained from   $\overline{h^i_j (\nk,\eta) h^j_i (\nk',\eta)   } $, where the bar above the expression represents an ensemble average over possible realizations, each one associated to the stochastic collapse. Therefore, from \eqref{tensormodescolapso}, the tensor power spectrum will involve 4-products of linear perturbations, e.g.  $\overline{ \bra \hat y (\nk_1,\eta)   \ket \bra \hat y (\nk_2,\eta)   \ket \bra \hat y (\nk_3,\eta')   \ket \bra \hat y (\nk_4,\eta')   \ket  } $ (recall that $\hat y \equiv a \hat \dphi$).  Thus, our model's prediction for the tensor power spectrum is (we invite the interested reader to consult Refs. \cite{lucila,lmosshort,lmosbig} for more technical details): } 
\begin{equation}
\mP_t (k) \simeq \epsilon_1^2 \mP_{s}^2,
\end{equation}
  {which implies an essentially undetectable amplitude of primordial B--modes. } In fact, our estimate for the tensor--to--scalar ratio is $r \simeq \epsilon_1^2 10^{-9}$. We can compare that expression with the standard one $r^{\std} = 16 \epsilon_1$. Thus, in our model  any inflationary potential results in a very small amplitude of tensor modes. Consequently, in our approach we can safely neglect the $r$ parameter in any data analysis.

We end this section by mentioning that we have proposed and analyzed other different collapse schemes. However, a Bayesian evidence analysis showed  a moderate  preference for \wig $ $ collapse  scheme model over the $\Lambda$CDM model using recent CMB and BAO data \cite{micol2}. For other collapse schemes the results of the statistical analysis showed preference for the $\Lambda$CDM model or inconclusive preference. That is the reason for choosing \wig $ $ collapse scheme in the present analysis.

\section{Inflationary potentials and the theoretical predictions}\label{potentials}

The standard analysis regarding the viability of inflationary potentials mainly relies on the scalar spectral index $n_s^\std$ and  tensor--to--scalar ratio $r^\std$.  From the predicted values in terms of slow roll parameters $n_s^\std = 1-2 \ei - \eii$, $r^\std= 16 \ei$, and given a particular potential $V$ with some parameter $\lambda$, it is then straightforward to write $n_s^\std$ and $r^\std$ as a function of the potential parameter $\lambda$ using  \eqref{PSR} (since slow roll parameters are not exactly constant, they have an extra dependence on the number of e-foldings). Then, observational constraints on $n_s$ and $r$ are compared with their predicted values as a function of $\lambda$. In this way, one explores a range of $\lambda$ values allowing us to accept (or not) the feasibility of a potential $V$ \cite{jmartin2013,jmartin2014}. 

In the preceding section we have seen that the predicted inflationary parameters given by \wig $ $ collapse scheme are different from standard inflation. In particular, the spectral index $n_s$ and tensor--to--scalar ratio $r$ are not related anymore (at first order in the slow roll parameters) because of the generic predicted smallness in $r$. Additionally, the spectral index as a function of slow roll parameters is also different from the traditional one, namely $n_s = 1+2 \ei - \eii$.

One may conclude that given the limits on $n_s$ established with \textit{Planck} data, the predicted $n_s$ given by \wig $ $ scheme must be within that interval. Nonetheless that conclusion assumes that the usual observational constraints on $n_s$ translate directly to our model. 
That is not the adequate procedure. In order to test our model's predictions, we must perform first a statistical analysis using recent observational data to obtain the constraints on $n_s$ in the context of the collapse models. In fact, as we will show in next section, the extra $k$ dependence in $\mP_s(k)$ introduced by the function $Q(z_k)$, see \eqref{Pk}, enlarges current observational bounds $n_s$ consistent with data. In the following, we provide the steps detailing the analysis process. 


\begin{itemize}
	
\item \textbf{Step 1}: Given the scalar power spectrum in Eq. \eqref{Pk}, we perform a statistical analysis using recent CMB and BAO data. The parameters we are interested in are: the \textit{Wigner} scheme parameters  $A,B$ and the inflationary parameters $n_s,A_s$. For a fixed set of $A$ values, we  find the posterior probability densities corresponding to the cosmological parameters, which include $A_s, n_s$ and the collapse parameter $B$.

\item  \textbf{Step 2}: For a given particular potential $V(\phi)$, with a single parameter ${\lambda}$, it is convenient  to calculate $\phi$ as a function of the number of e-folds from the ``horizon crossing'' [that is from the time $\eta_*$ such that the pivot scale $k_*$ satisfies $k_*=a(\eta_*) H(\eta_*)$] to the end of inflation; we denote such period of e-foldings as $\DN$. 

 Therefore, one needs to solve the equation of motion for the homogeneous part of the field $\phi_0(\eta)$ in the slow roll approximation together with Friedmann's equation. That is, after a change of variables $N(\eta)$, where $N$ is the number of e-foldings from the beginning of inflation to some time $\eta$, the equations to solve are: $3 \mH^2 \simeq a^2 V/M_P^2$ and $ 3 \mH^2 \frac{d \phi_0}{d N} \simeq -a^2 \partial_\phi V$. Those equations can be combined to yield
\begin{equation}\label{eqxx}
\frac{d \phi}{d N} = -M_P^2 \frac{d \ln V}{d \phi},
\end{equation}
note that for ease of notation we have omitted the subindex 0 from the background field $\phi_0$. Denoting by $\mI$ the primitive
	\begin{equation}\label{prim}
	\mI_{\lambda} (\phi) \equiv \int^\phi d\varphi \frac{V_{\lambda}(\varphi)}{\partial_\phi V_{\lambda} (\varphi)},
	\end{equation}
equation \eqref{eqxx} can be solved
\begin{equation}
N =  - \frac{1}{M_P^2} [ \mI_{\lambda}(\phi ) - \mI_{\lambda}(\phi_{\ini}) ]. 
\end{equation}	
Therefore one has
\begin{subequations}
	\begin{equation}
	N_{\fin} = - \frac{1}{M_P^2} [ \mI_{\lambda}(\phi_{\fin} ) - \mI_{\lambda}(\phi_{\ini}) ],
	\end{equation}
	\begin{equation}
	  N_* = - \frac{1}{M_P^2} [ \mI_{\lambda}(\phi_*) - \mI_{\lambda}(\phi_{\ini}) ], 
	\end{equation}
\end{subequations}	
	formally $\phi_*$ represents the vacuum expectation value of the field evaluated when the pivot scale crosses the Hubble radius; $N_{\fin}$, represents the number of e-folds from the beginning to end of inflation; and $N_*$ represents the number e-folds from the beginning of inflation to the conformal time $\eta_*$ where the pivot scale $k_*$ crossed the horizon.     From the latter expressions, it follows that
	\begin{equation}\label{phi*}
	\phi_* = \mI_{\lambda}^{-1} [ \mI_{\lambda}(\phi_{\fin}) + M_P^2 \DN     ]
	\end{equation}
where $\DN \equiv N_{\fin} - N_*$.	Equation \eqref{phi*}, explicitly yields $\phi_*$ as a function of the potential's parameter $\lambda$ and $\DN$.

\item  \textbf{Step 3} For the particular potential given $V_\lambda (\phi)$, we express the slow roll parameters in terms of $\lambda$ and $\DN$, i.e. we need to find the explicit expressions: $\ei(\lambda,\DN)$, $\eii(\lambda,\DN)$. 

In order to achieve that, we first use the definition of the slow roll parameters in terms of the potential $V$ and its derivatives $\partial_\phi V$ and $\partial_{\phi \phi} V$, eqs. \eqref{PSR}. After inserting the explicit form of the potential evaluated at $\phi_*$, i.e. $V_\lambda (\phi_*)$, into eqs. \eqref{PSR}, that operation yields $\ei(\lambda,\phi_*)$, $\eii(\lambda,\phi_*)$. Finally, we use solution  \eqref{phi*} to obtain $\ei(\lambda,\DN)$, $\eii(\lambda,\DN)$. Recall that we can find the value of the field at the end of inflation by using the condition $\ei(\phi_{\fin}) \simeq \eii(\phi_{\fin}) \simeq 1$.

\item  \textbf{Step 4} With the expression for the slow roll parameters at hand $\ei(\lambda,\DN)$, $\eii(\lambda,\DN)$, we write $n_s$ also as a function of $\lambda$ and $\DN$, i.e. 
   \begin{equation}
    n_s(\lambda,\DN) = 1+2\ei(\lambda,\DN) - \eii(\lambda,\DN)
   \end{equation}
We remind the reader that our model's prediction for $r$ is extremely small, so we neglect it from the analysis.

\item  \textbf{Step 5}: For distinct values of $\DN$ and $\lambda$ we obtain different predicted values of $n_s(\lambda,\DN)$. Therefore, we can compare our predicted values with the ones obtained in \textit{Step 1} from the data analysis. Since our predicted $n_s$ and $r$ are generically different from the usual one, we expect a difference in the type of inflationary potentials that are allowed between our approach and the standard one. 
	
\end{itemize}

As shown in Refs. \cite{jmartinreheating,jmartinreheating2}, in \emph{Step 5} we need to take into account the reheating era in order to choose  a $\DN$ that is physically possible. One can define the reheating parameter as
\begin{equation}
\ln R_{\rad} = \frac{1-3 \bar{w}_{\reh}}{12(1+\bar{w}_{\reh})} \ln \left( \frac{\rho_{\reh}}{\rho_{\fin}} \right)
\end{equation}
where $\bar{w}_{\reh}$ is the mean equation of state parameter during reheating, $\rho_{\reh}$ is the energy density at the end of the reheating era, and $\rho_{\fin}$ is the energy density at the end of inflation. Consequently, it has been shown \cite{jmartinreheating,jmartinreheating2} that the quantities $\DN$ and $R_{\rad}$ are related by
\begin{equation}\label{masterreh}
\DN = \ln R_{\rad} - N_0 -\frac{1}{4} \ln \left[ \frac{9}{2 \epsilon_{1*}} \frac{V_{\fin}}{V_*} \right] + \frac{1}{4} \ln (8 \pi^2 A_s)
\end{equation}
with
\begin{equation}
N_0 \equiv \ln \left( \frac{k_*/a_0}{ \rho_\gamma^{1/4}} \right);
\end{equation}
the quantity $ \rho_\gamma$ denotes the energy density of radiation today, and $k_*/a_0$ the pivot scale normalized at the scale factor today. Taking the pivot scale $k_*/a_0 = 0.05$ Mpc$^{-1}$ and recent bounds on $\rho_\gamma$ \cite{planck2015} implies that $N_0 \simeq -61.7$.

From the definition of the reheating parameter, we see that $\ln R_{\rad}$ is not arbitrary since $-1/3 < \bar{w}_{\reh} < 1$ and $\rho_{\nuc} < \rho_{\reh} < \rho_{\fin}$. Consequently, the quantity $\DN$ is also constrained to vary in the range $\DN \in [\DN^{\nuc},\DN^{\fin} ]$. That is, $\DN^{\fin}$ corresponds to assume $\rho_{\reh} = \rho_{\fin}$ which means that reheating takes place instantaneously after inflation ends. And $\DN^{\nuc}$ corresponds to assume that $\rho_{\reh} = \rho_{\nuc}$, i.e. that the reheating era extends up to the nucleosynthesis  epoch. Moreover, that range is model-dependent since $\rho_{\fin}$ or $V_{\fin}/V_*$ differ for different inflationary scenarios. It is shown that  $\DN^{\nuc}$ and $\DN^{\fin}$ are given by \cite{jmartinreheating,jmartinreheating2}
\begin{eqnarray}\label{Nnuc}
\DN^{\nuc} &=& -N_0 + \ln \left( \frac{H_*}{M_P} \right) - \frac{1}{3(1+\bar w_{\reh})} \ln \frac{\rho_{\fin}}{M_P^4} \nn
& +& \frac{1-3\bar w_{\reh}}{12(1+\bar w_{\reh})} \ln \frac{\rho_{\nuc}}{M_P^4},
\end{eqnarray}
with  $\rho_{\nuc} \simeq (10$ MeV$  )^{4}$,
and
\begin{equation}\label{Nend}
\DN^{\fin} = -N_0 + \ln \left( \frac{H_*}{M_P} \right) - \frac{1}{4} \ln \frac{\rho_{\fin}}{M_P^4}.
\end{equation}
Note that these equations are algebraic for $\DN$ since $H_*$ and $\rho_{\fin}$ depend on $\DN$.

Fortunately enough, with help of the ASPIC code, we automatize \textit{Steps 2, 3, 4} and \textit{5} for each inflationary potential. Additionally, in order to be able to compare our results and the standard ones, we repeat all steps but considering the traditional predictions from inflation, that is, we consider the usual $n_s^\std$ and $r^\std$ (so in \textit{Step 4} we use $n_s^\std = 1-2 \ei - \eii$, $r^\std= 16 \ei$ ).

In Table \ref{listapotenciales}, we show a list of the inflationary potentials considered in our analysis, depicting their specific shape and characteristic parameter. The potentials correspond to popular inflationary models found in literature, these models are: Higgs Inflation (HI) also known as Starobinsky's or $R^2$ inflation; Large Field Inflation (LFI),  Radiatively Corrected Massive Inflation (RCMI), Radiatively Corrected Quartic Inflation (RCQI), Radiatively Corrected Higgs Inflation (RCHI), Natural Inflation (NI), Exponential SUSY Inflation (ESI), Power Law Inflation (PLI), Double Well Inflation (DWI) and Loop Inflation (LI). For the theoretical motivation of these models we refer the reader to  Ref. \cite{jmartin2013} (and references therein), where a brief review of each model is given.

All the potentials we have chosen for our analysis contain a ``mass" term $M$ associated to the characteristic energy scale of inflation, which in turn is related to the spectra's amplitude. However, we do not treat $M$ as a potential's parameter. As a matter of fact, the first potential on the list corresponds to Higgs Inflation (HI) which only contains $M$ and no other parameter, hence HI is parameterless. All the rest of potentials considered contain a single parameter.

\begin{table}[]
	\centering
	\caption{List of analyzed inflationary potentials }
	\label{listapotenciales}
	\begin{tabular}{llc}
		\hline
		\multicolumn{1}{c}{Inf. Model} & \multicolumn{1}{c}{Potential $V(\phi)$} & 
		Parameter  \\ \hline
		HI & $ M^4 (1-e^{-\sqrt{\frac{2}{3}}\frac{\phi}{M_P}   })^2 $ & NA  \\ 
		LFI  & $ M^4 (\phi/M_P)^p$           & $p$   \\ 
		RCQI   &   $ M^4 (\frac{\phi}{M_P})^4 [1-\alpha \ln (\frac{\phi}{M_P})]$    &   $ \alpha$  \\ 
		RCHI & $ M^4 [1-2 e^{-\frac{2\phi}{\sqrt{6}M_P}} + \frac{A_I}{16\pi^2} 	\frac{\phi}{\sqrt{6} M_P} ]$ & $A_I$ \\ 
		RCMI & $ M^4 (\frac{\phi}{M_P})^2 [1-2\alpha (\frac{\phi}{M_P})^2  \ln 	(\frac{\phi}{M_P}) ] $ & $\alpha$   \\ 
		NI  & $M^4 [1 + \cos (\frac{\phi}{f})] $ & $f$  \\ 
		ESI & $ M^4 (1-e^{-q\phi/M_P})$ & $q$  \\ 
		PLI & $ M^4 e^{-\alpha \phi/M_P}$ & $\alpha$  \\ 
		DWI & $ M^4 [(\frac{\phi}{\phi_0}  )^2 -1 ]^2$ & $\phi_0$  \\ 
		LI & $M^4 [1+\alpha \ln(\frac{\phi}{M_P})  ]   $ & $\alpha$  \\ \hline
	\end{tabular}
\end{table}

\section{Results and Discussion}\label{results}

For \textit{Step 1}, we used a modified version of the CAMB \cite{CAMB} code to include the primordial power spectrum of the \textit{Wigner's} scheme. Therefore, we considered an extension of the minimal $\Lambda$CDM model, adding the collapse parameters $A$ and $B$ to the usual set of cosmological parameters: the baryon density $\Omega_b h^2$, the cold dark matter density $\Omega_c h^2 $, the ratio between the sound horizon and the angular diameter distance at decoupling $\theta$, the optical depth $\tau$, the primordial scalar amplitude $A_s$ and the scalar spectral index $n_s$. Concerning the data analysis we work with flat priors for the cosmological parameters.
We use the CMB anisotropy and polarization spectrum reported by the  Planck Collaboration ~\cite{planck2015} together with data from Baryonic Acoustic Oscillation (BAO). In particular, we consider the high-$\ell$ Planck temperature data from the 100-,143-, and 217-GHz half-mission T maps, and  the low-$\ell$ data by the joint TT, EE, BB and TE likelihood. Also, we consider BAO data by the 6dF Galaxy Survey (6dFGS)~\cite{Beutler:2011hx}, SDSS DR7 Main Galaxy Sample (SDSS-MGS) galaxies~\cite{Ross:2014qpa},  BOSS galaxy samples, LOWZ and CMASS~\cite{Anderson:2013zyy}.   

It follows from \ref{Pk} and \ref{etac} that the scalar spectrum amplitude $A_s$ is degenerated with  the  collapse $A$ parameter. Therefore, we test several values of $A$ by taking $B=0$.  We analyze the full interval $-10^{8} \leq A \leq -10^{2}$ and divide it in subintervals of 1 order of magnitude. For all $A$ in such an interval, the condition $-k\tc \gg 1$ is satisfied, with $k$ between $10^{-6}$ Mpc$^{-1}$  and $10^{-1}$ Mpc$^{-1}$. Moreover, for each subinterval $-10^{i+1} \leq A \leq -10^{i}$ (with $i$ an integer such that $i \in [2,7]$), we select the value of $A$ which minimizes the variation of $A_s$ with respect to the value obtained in the statistical analysis with observational data for the  standard $\Lambda$CDM model. 

For each fixed value of $A$ chosen in the aforementioned manner, we now include the $B$ parameter as a free parameter in our analysis along with the rest of cosmological parameters. We perform a Monte Carlo Markov chain analysis using the available package COSMOMC \cite{COSMOMC}.

\begin{figure}
	\begin{center}
		\includegraphics[scale=0.95]{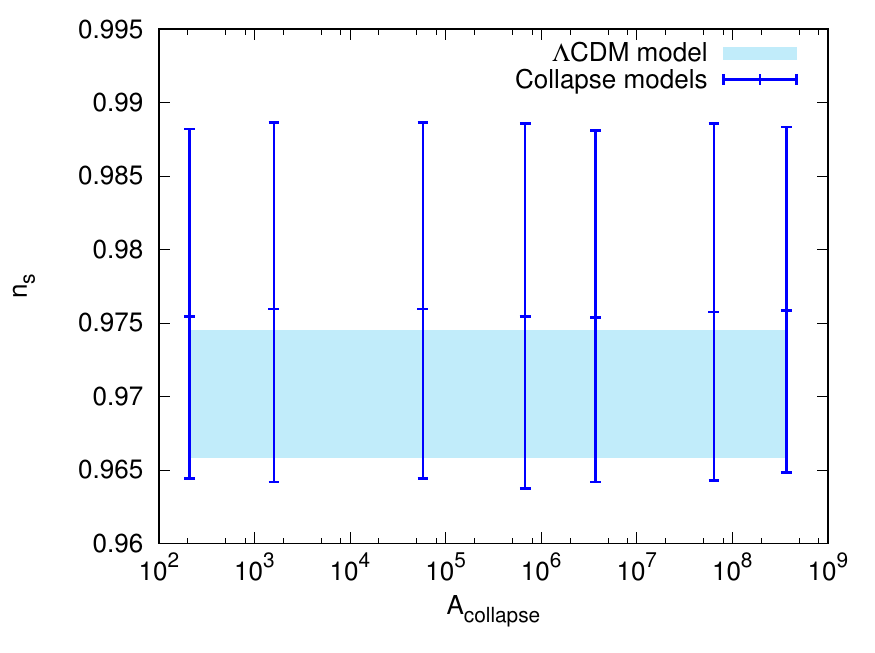}
	\end{center}
	\caption{Estimated 68\% confidence limits on $n_s$  for distinct fixed values of $A$ using \textit{Planck 2015} + BAO data set. The $A$ parameter is negative. The shadowed area corresponds to the 68 \% confidence limits of $n_s$  for the $\Lambda$CDM model using the same data set.}
	\label{fig1}
\end{figure}

\begin{figure}
	\begin{center}
		\includegraphics[scale=0.8]{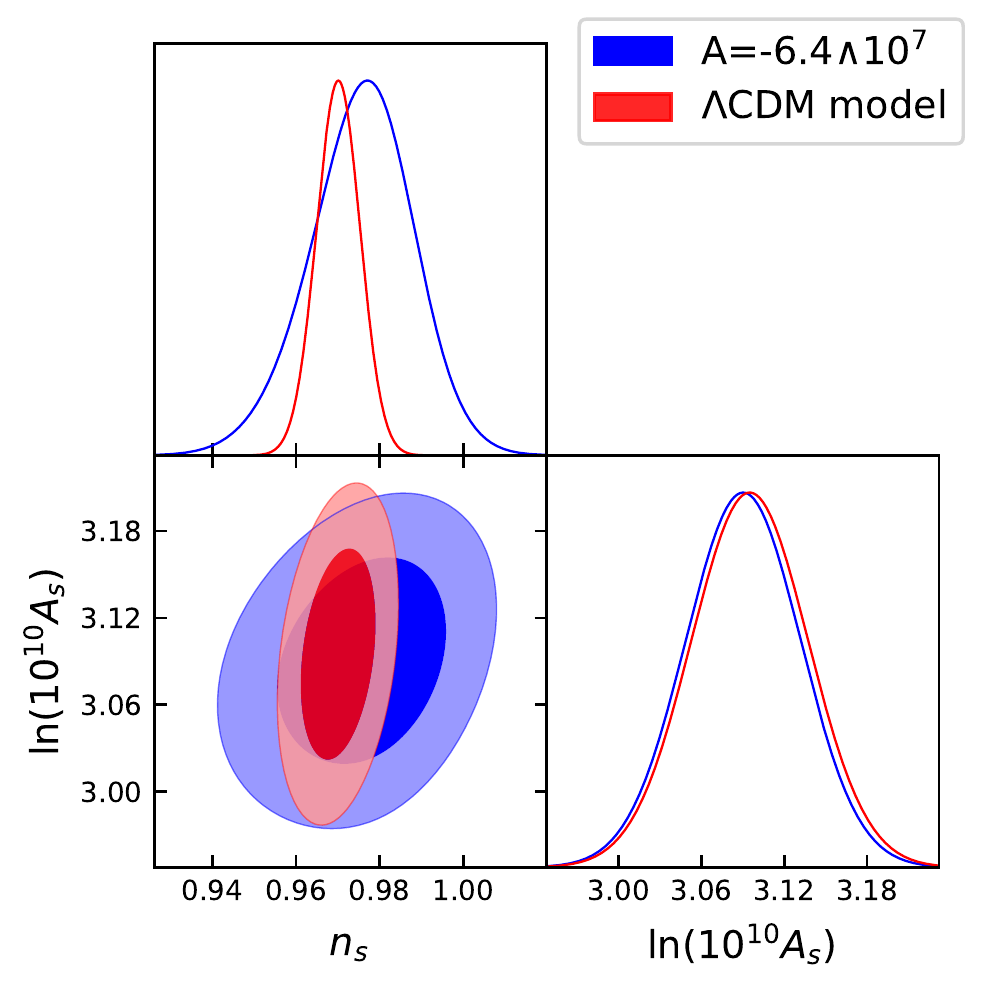}
		\includegraphics[scale=0.8]{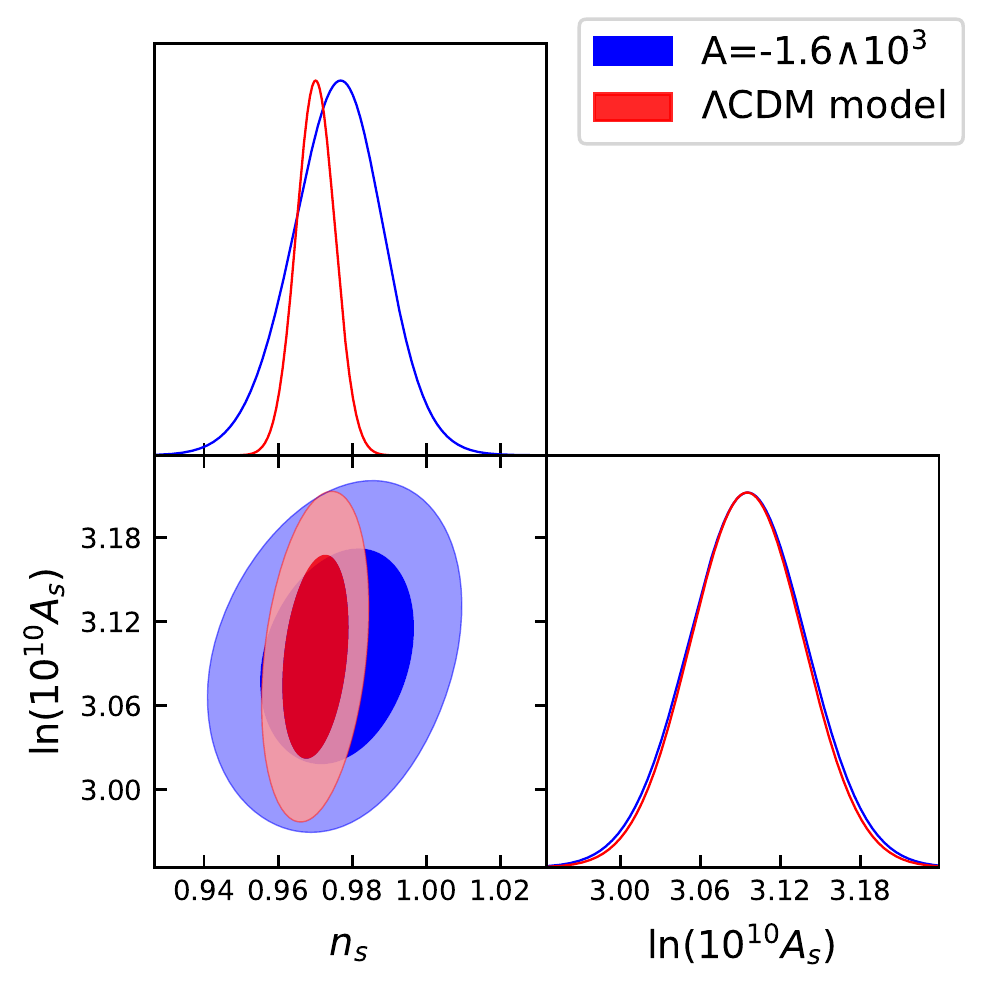}
	\end{center}
	\caption{68\% and 95\% two dimensional confidence regions and posterior probability density of  $n_s$ and $A_s$ for two values of $A$ using Planck 2015 + BAO data set. We also show the same confidence regions  of the $\Lambda$CDM model (red) in the two plots. Other values of $A$ exhibit the same behavior.   }
	\label{fig3}
\end{figure}

In Fig. \ref{fig1}, we show  the 68\% confidence limits on  $n_s$  from the data analysis obtained for each value of $A$.  For comparison purposes, we also include the respective limits on $n_s$ for the minimal $\Lambda$CDM model. Since the modification of the standard cosmological model that we are studying in this paper involves only a change in the standard inflationary model, we will refer to it in what follows as the standard framework or standard inflationary model.
 It follows from Fig. \ref{fig1}  that the estimated confidence intervals are  very similar for all the values of $A$ selected previously, which covers an interval of several orders of magnitude. In Fig. \ref{fig3}, we show the 1$\sigma$ and 2$\sigma$ confidence regions of the inflationary parameters $n_s$ and $A_s$ for two values of $A$; other values of $A$ result in similar plots. Also, we include the same confidence regions for the standard $\Lambda$CDM model. As we can observe, the effect of including \wig $ $ collapse scheme is to enlarge the estimated interval of $n_s$ with respect to the standard model.  Furthermore, the analysis above indicates that the estimated value of $n_s$ is independent of the specific value associated to the time of collapse. As we will see next, that result, together with the generically predicted smallness of the $r$ parameter, will modify the usual conclusions regarding the type of potentials allowed by observations.

Following \textit{Steps 1--5}, we analyze the viability of all inflationary potentials listed in Table \ref{listapotenciales}. In order to provide a more detailed picture of the analysis made, we focus on three particular potentials that serve as an example: LFI,  PLI and RCQI. 

\begin{figure}
	\begin{center}
		\includegraphics[scale=1.0]{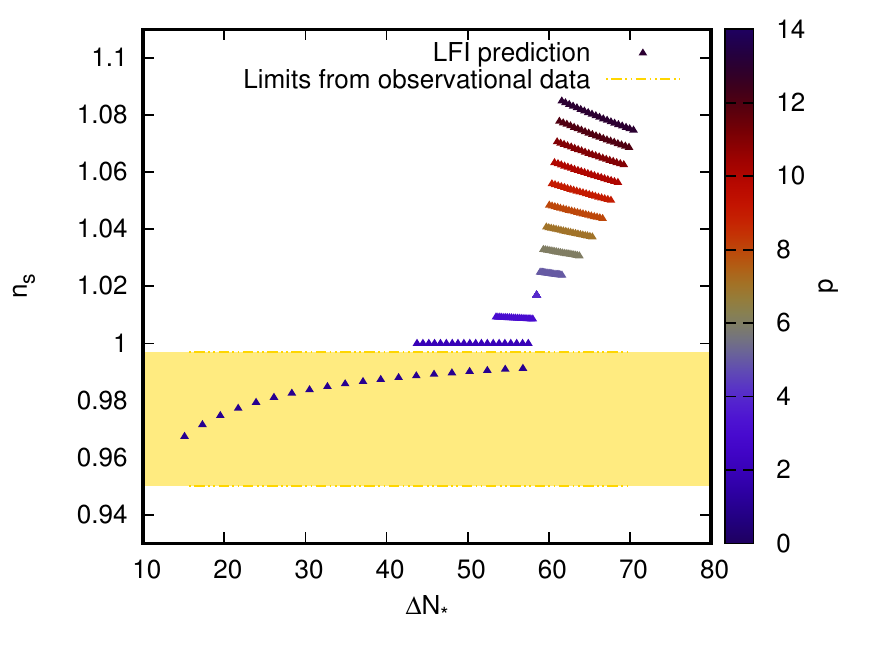}
	\end{center}
	\caption{The predicted spectral index $n_s$ as a function of $\DN$ with varying parameter $p$  for the LFI potential in \wig $ $ scheme. The shadowed region corresponds to the estimated $n_s$ at 2$\sigma$ CL using \textit{Planck 2015} + BAO data set. }
	\label{LFInscol}
\end{figure}

\begin{figure}
	\begin{center}
		\includegraphics[scale=1.0]{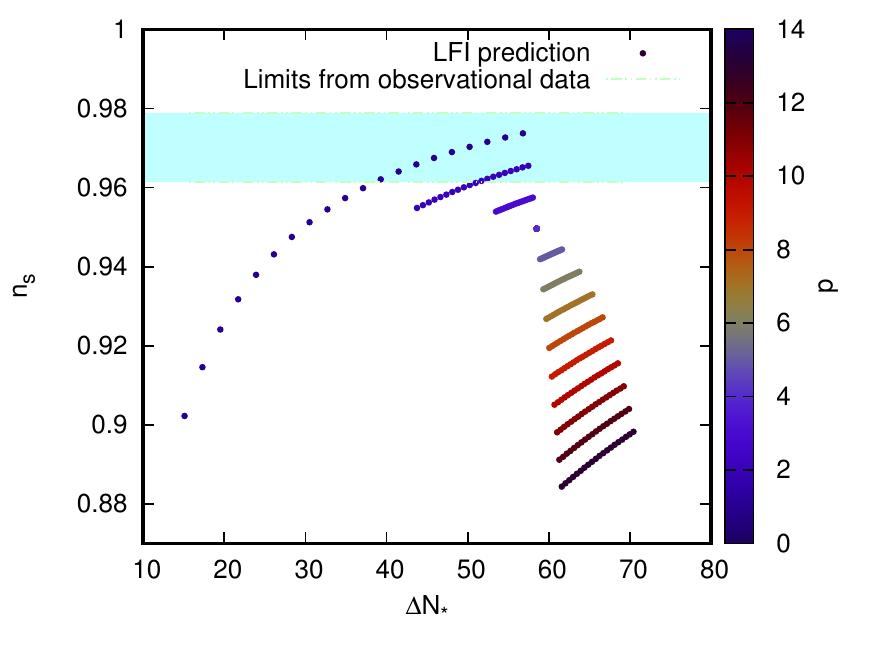}
	\end{center}
	\caption{The predicted spectral index $n_s$ as a function of $\DN$ with varying parameter $p$  for the LFI potential in the standard framework. The shadowed region corresponds to the estimated $n_s$ at 2$\sigma$ CL using \textit{Planck 2015} + BAO data set.}
	\label{LFInsstd}
\end{figure}

\begin{figure}
	\begin{center}
		\includegraphics[scale=1.0]{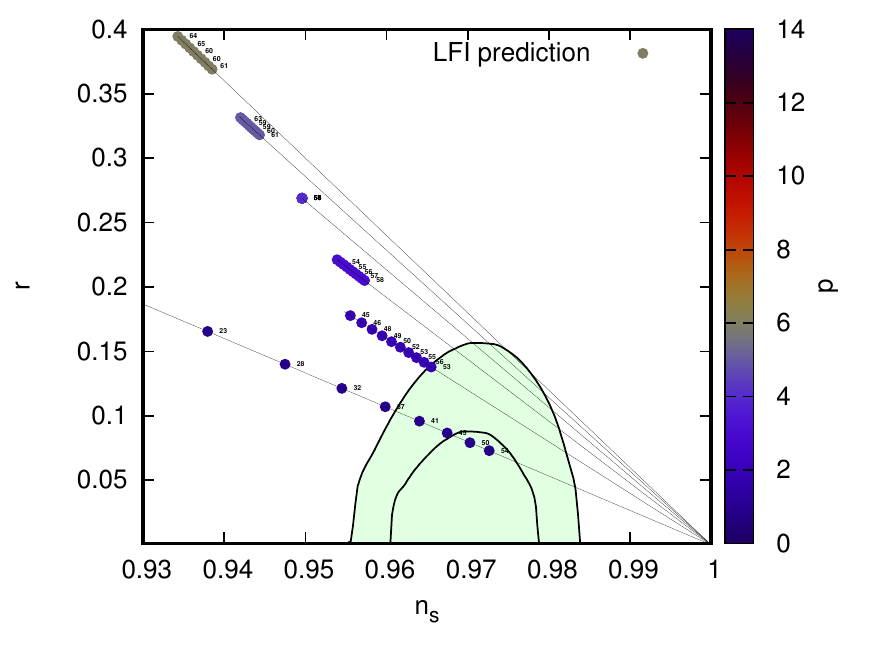}
	\end{center}
	\caption{The predicted $n_s$ and $r$ as a function of $\DN$ with varying parameter $p$  for the LFI potential in the standard framework. The shadowed region corresponds to the marginalized 68\% and 95\% CL regions for $n_s$ and $r$ from \textit{Planck 2015} + BAO data set. }
	\label{LFInsr}
\end{figure}

 Let us begin by analyzing the LFI potential with varying parameter $p$. In Figs. \ref{LFInscol} and \ref{LFInsstd} we  show the predicted  values of $n_s$ as a function of $\DN$ for \wig $ $ collapse scheme and  the standard framework respectively. In each figure we also include the 2$\sigma$ confidence interval  resulting from the statistical analysis using recent CMB and BAO data. As we observe from the figures, the predicted interval for the standard framework   is different from the respective one in \wig $ $ scheme. In the standard inflationary model we see that for $p \in [1,2]$, the predicted values of  $n_s$ for some particular values of $\DN$  lie within the $2\sigma$ confidence interval allowed by the data set. Moreover,  it follows from Fig. \ref{LFInsr} that including the $r$ parameter in the analysis restricts the viability of   the standard framework  to a smaller interval of $\DN$. Meanwhile, in \wig $ $ scheme, we observe from Fig. \ref{LFInscol}  that the LFI potential is only viable for $p \simeq 1$ and some particular values of $\DN$. Let us recall that in  \wig $ $ scheme we do not include in the analysis the $r$ parameter since the model generically predicts a strong suppression of primordial tensor modes.  In brief, the allowed values of the free parameter $p$ and $\DN$ for the \textit{Wigner} collapse scheme  are different from the respective ones for the standard inflationary model.

\begin{figure}
	\begin{center}
		\includegraphics[scale=1.0]{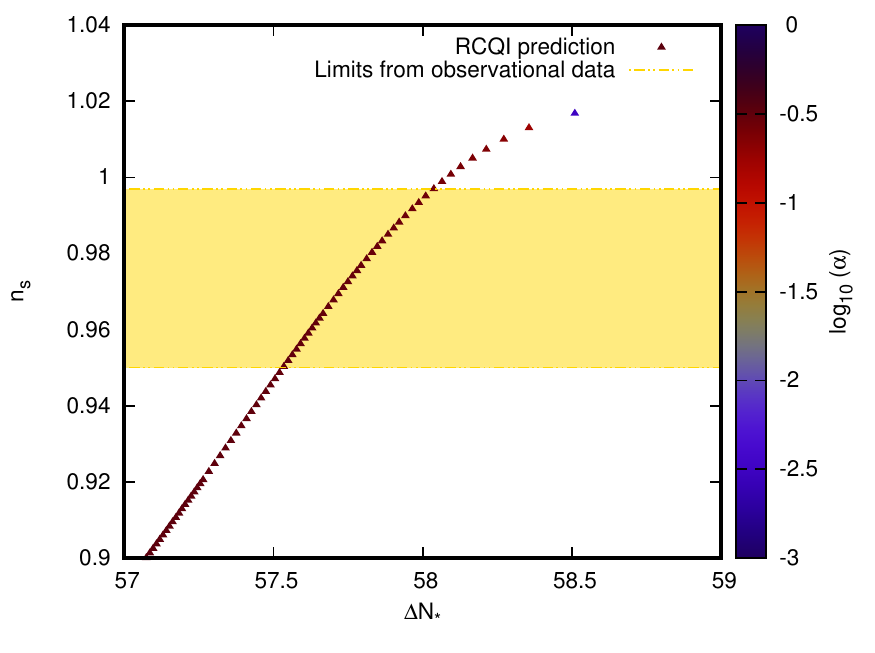}
	\end{center}
	\caption{The predicted spectral index $n_s$ as a function of $\DN$ with varying parameter $\alpha$  for the RCQI potential with $\omega_{reh}= - \frac{1}{3}$ in \wig $ $ scheme. The shadowed region corresponds to the estimated $n_s$ at 2$\sigma$ CL using \textit{Planck 2015} + BAO data set. }
	\label{RCQInscol}

\end{figure}

\begin{figure}
	\begin{center}
		\includegraphics[scale=1.0]{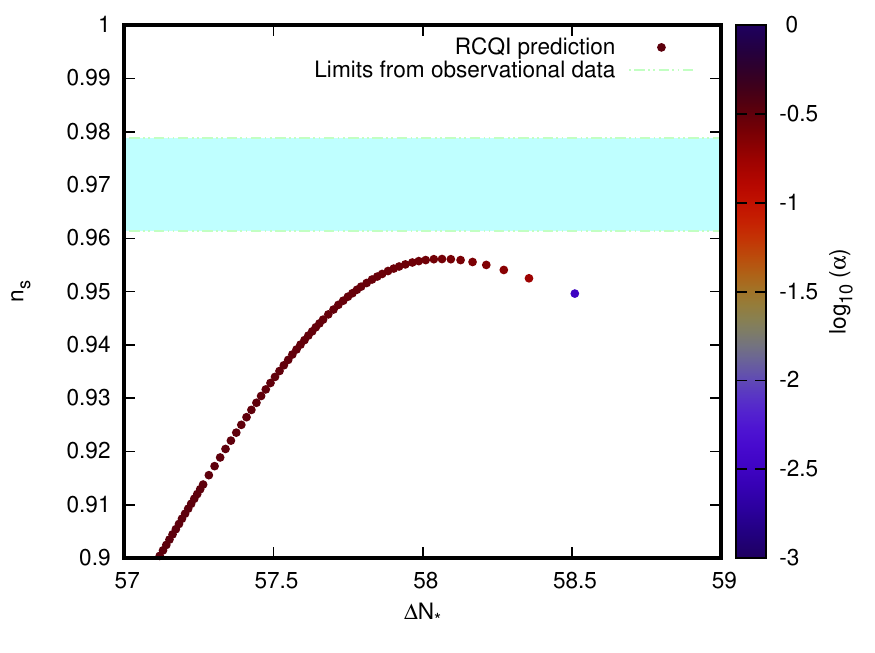}
	\end{center}
	\caption{The predicted spectral index $n_s$ as a function of $\DN$ with varying parameter $f$  for the RCQI potential with $\omega_{reh}= - \frac{1}{3}$ in the standard framework. The shadowed region corresponds to the estimated $n_s$ at 2$\sigma$ CL using \textit{Planck 2015} + BAO data set. }
	\label{RCQInsstand}
\end{figure}

\begin{figure}
	\begin{center}
		\includegraphics[scale=1.0]{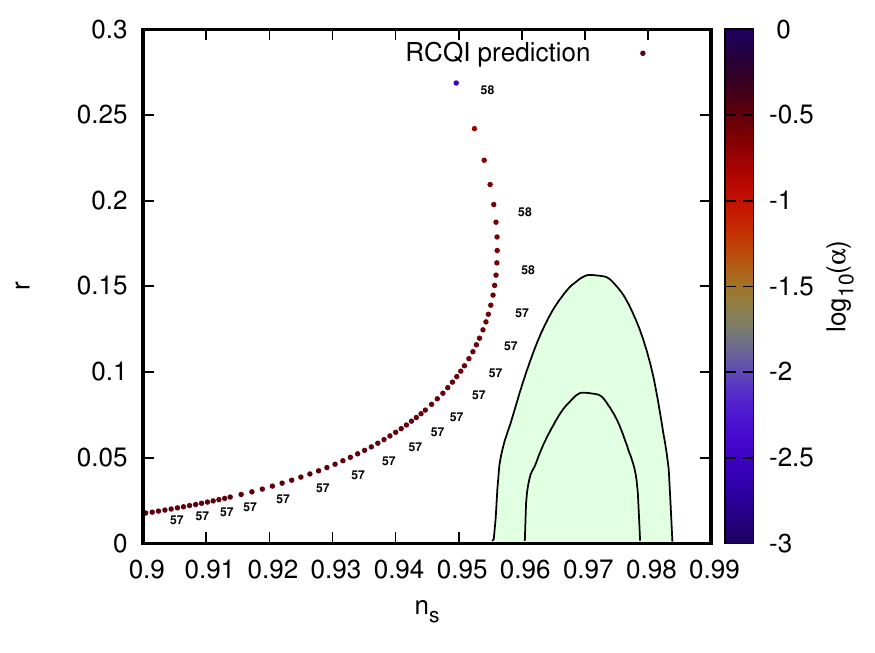}
	\end{center}
	\caption{The predicted $n_s$ and $r$ as a function of $\DN$ with varying parameter $\alpha$ for the RCQI potential with $\omega_{reh}= - \frac{1}{3}$ in the standard framework. The shadowed region corresponds to the marginalized 68\% and 95\% CL regions for $n_s$ and $r$ from \textit{Planck 2015} + BAO data set. }
	\label{RCQInsr}
\end{figure}

Another interesting potential to use as an example is the RCQI model with $\omega_{reh}= - \frac{1}{3}$.  Fig. \ref{RCQInscol}  shows
the predicted  values of $n_s$ as a function of $\DN$ in \wig $ $ collapse scheme together with the $2\sigma$ confidence interval determined from the statistical analysis using CMB and BAO data. We observe that the predicted value of $n_s$ for some values of the free parameters $\log \alpha$ and $\DN$ lie in the region allowed by the observational data. On the other hand, it follows from Figs. \ref{RCQInsstand} and \ref{RCQInsr} that all predicted values of $n_s$ and $r$  in the context of  the standard framework lie outside the allowed region by the data set. In summary, the RCQI potential is ruled out by the data in the context of the standard inflationary model while for the \textit{Wigner} collapse scheme there is a set of values of the free parameter $\log \alpha$ and $\DN$ for which the prediction of $n_s$ is in agreement with the observational data.

\begin{figure}
	\begin{center}
		\includegraphics[scale=1.0]{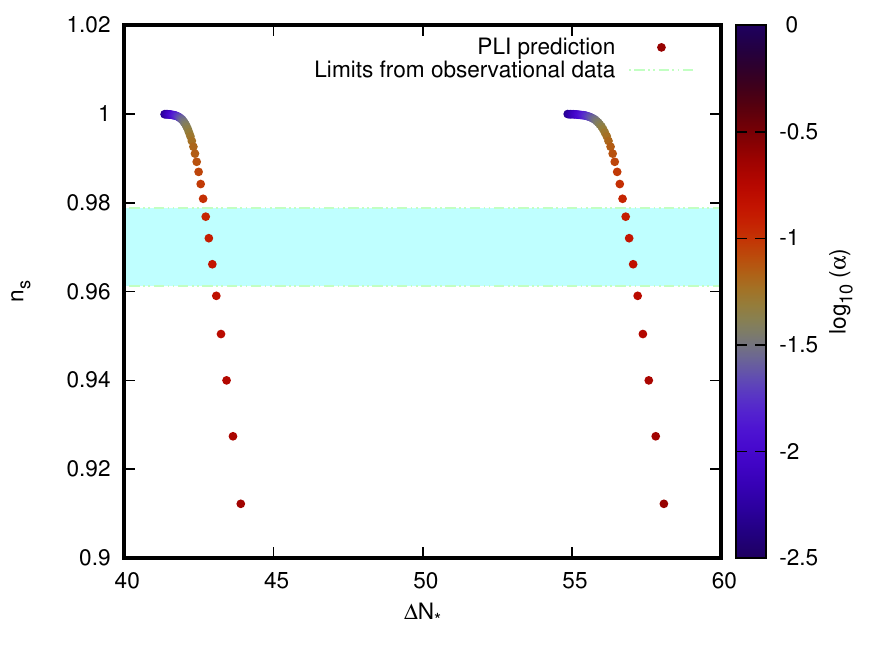}
	\end{center}
	\caption{The predicted spectral index $n_s$ as a function of $\DN$ with varying parameter $\alpha$  for the PLI potential in the standard framework. The shadowed region corresponds to the estimated $n_s$ at 2$\sigma$ CL using \textit{Planck 2015} + BAO data set. }
	\label{PLInsstand}
\end{figure}

\begin{figure}
	\begin{center}
		\includegraphics[scale=1.0]{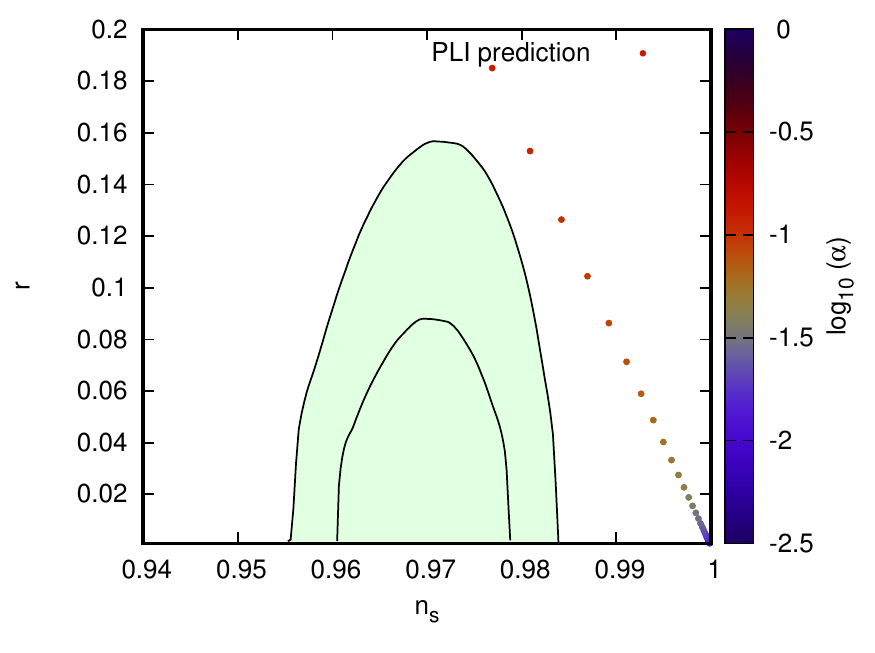}
	\end{center}
	\caption{The predicted $n_s$ and $r$ as a function of $\DN$ with varying parameter $\alpha$ for the PLI potential in the standard framework. The shadowed region corresponds to the marginalized 68\% and 95\% CL regions for $n_s$ and $r$ from \textit{Planck 2015} + BAO data set. }
	\label{PLInsr}
\end{figure}

\begin{figure}
	\begin{center}
		\includegraphics[scale=1.0]{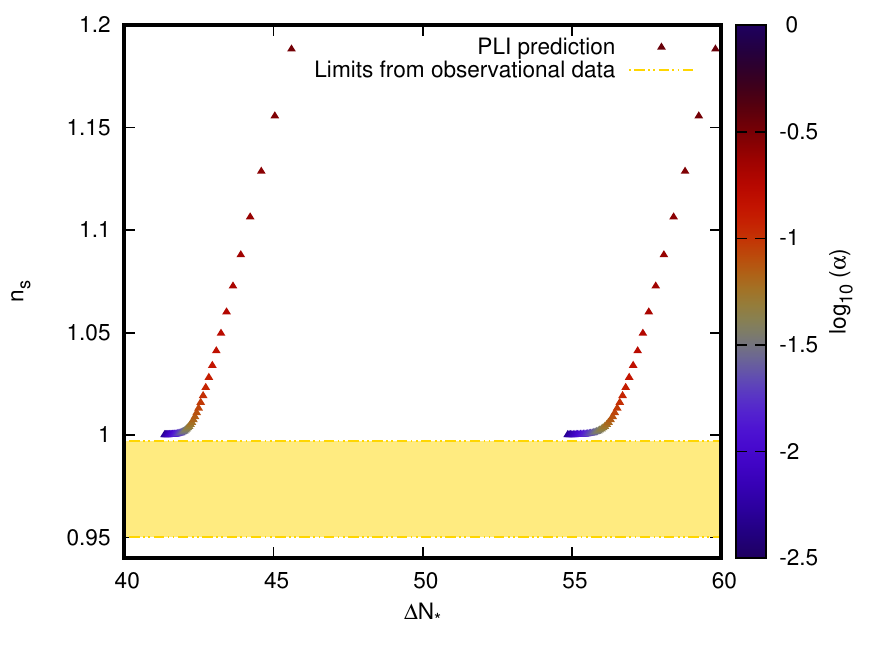}
	\end{center}
	\caption{The predicted spectral index $n_s$ as a function of $\DN$ with varying parameter $\alpha$ for the PLI potential in \wig $ $ scheme. The shadowed region corresponds to the estimated $n_s$ at 2$\sigma$ CL using \textit{Planck 2015} + BAO data set. }
	\label{PLInscol}
\end{figure}

We choose as a final example of our analysis the historical potential given by PLI with varying parameter $\alpha$. For the standard inflationary  model, we observe from Figs.  \ref{PLInsstand}  and \ref{PLInsr}  that there is no value of $\alpha$ and $\DN$ that makes the predicted $n_s$ and $r$ to lie inside the 2$\sigma$ confidence region in the $n_s$--$r$ plane. Furthermore, Fig. \ref{PLInscol} shows that there is no value of the $\alpha$ parameter for which the prediction of $n_s$ lies in the allowed region by the data in the collapse framework.
Thus, we conclude that the PLI potential is not viable for both the \textit{Wigner} $ $ scheme and the standard framework.

The results obtained for all the potentials considered are summarized in Tables \ref{tabla2} and \ref{tabla3}. In Table \ref{tabla2}, we indicate for each inflationary potential considered in this paper the values of the potential's parameter and $\DN$ such that the predicted  value of $n_s$ for \wig $ $ collapse scheme lies inside the estimated 1$\sigma$ and/or 2$\sigma$ confidence interval obtained from the statistical analysis with recent CMB and BAO data. In Table \ref{tabla3},  for the standard inflationary model, we indicate for each inflationary potential the values of the potential's parameter and $\DN$ such that the predicted  value of the pair $n_s-r$, lie inside the estimated 1$\sigma$ and/or 2$\sigma$ confidence regions obtained from the CMB/BAO data analysis. We can distinguish different type of situations: (i) Potentials that are ruled out both in the collapse and in the standard inflationary model context: this is the case of the PLI model discussed above; (ii) Potentials that are ruled out in the context of the standard framework, but not in \wig $ $ scheme: here we refer to the RCQI potential for both cases of $\omega_{reh}$ considered in this paper; (iii) Potentials whose predictions are in agreement with the data but this happens for  different values of the free parameters in the context of the collapse models and the standard model: this is the case of NI, LFI, DWI, RCMI models and finally; (iv) similar to (iii) but the difference in the allowed range of the free parameters is small: here we refer to the HI, RCHI, LI with $\alpha < 0$ and ESI for both values of $\omega_{reh}$ considered in this paper. In summary, in this section we have shown, that the analysis of viable potentials for inflation in the context of \wig $ $ collapse model  is different from the one performed for the standard inflationary model in most of the cases studied in this work. 

\begin{table*}
	\centering
	\caption{Results:  The first column details  the inflationary potential considered; the second column refers to the  potential's parameter; the third column reports the parameter  values for \wig $ $ collapse scheme that are not discarded by recent CMB and BAO data; the fourth column reports  the $\DN$ values for \wig $ $ collapse scheme that are not discarded by recent CMB and BAO data; the fifth column refers to the confidence level for which the values reported in columns 3,4 are in agreement with recent CMB and BAO data.
	}
	\label{tabla2}
	%
	\begin{tabular*}{1.0\textwidth}{c @{\extracolsep{\fill}}cccc}\hline
		\multicolumn{1}{|c}{\textbf{Inf. model} }&
		\multicolumn{1}{|c}{\textbf{Parameter}}&
		\multicolumn{1}{|c}{\textbf{Param. values}}&
		\multicolumn{1}{|c}{\textbf{$\Delta N_*$}}&
		\multicolumn{1}{|c|}{\textbf{Conf. region}}
		\\
		\hline
		\\
		\vspace{0.1cm}	HI & - & - & $[42.10,55.68]$   & 2$\sigma$-1$\sigma$   \\ \hline
		\vspace{0.1cm}		LFI & $p$ & 1 & $[15.10,56.81]$   & 2$\sigma$-1$\sigma$   \\ \hline
		\vspace{0.1cm}		RCQI ($\bar{\omega}_{\text{reh}} = 0$) & $\log \alpha$  & -0.55, -0.50 &  $[44.36,56.49]$  & 2$\sigma$-1$\sigma$  \\ \hline
		\vspace{0.1cm}		RCQI ($\bar{\omega}_{\text{reh}} = -1/3$) & $\log \alpha$  & $[-0.57,-0.51]$ & $[57.53,58.03]$  & 2$\sigma$-1$\sigma$  \\ \hline
		RCHI & $A_I$ & $[-5.86,100]$  & $[41.96,56.47]$    & 2$\sigma$-1$\sigma$  \\ \hline
		RCMI & $\log \alpha$ & -6,-4.3,-4,-3.7 & $[43.46,56.71]$  & 2$\sigma$-1$\sigma$   \\ \hline
		\vspace{0.1cm}			NI & $\log (f/M_P)$ & 0.69,0.85,1  & $[43.54,57.49]$ & 2$\sigma$-1$\sigma$   \\ \hline
		\vspace{0.1cm}			ESI ($\bar{\omega}_{\text{reh}} = 0$ ) & $\log q$ & \makecell{-3,-1.3,-1,-0.6 \\ -0.3,0,0.18,0.54}  & $[41.35,56.81]$   & 2$\sigma$-1$\sigma$   \\ \hline
		\vspace{0.1cm}			ESI ($\bar{\omega}_{\text{reh}} = -1/3$)  & $\log q$ & \makecell{-3,-1.3,-1,-0.6 \\ -0.3,0,0.18,0.54} & $[30.41,56.81]$   & 2$\sigma$-1$\sigma$   \\ \hline
		\vspace{0.1cm}			PLI & $\log \alpha$ & None  & None  & None  \\ \hline 
		\vspace{0.1cm}			DWI &  $\log (\phi_0/M_P)$ & $[1.18,3]$  & $[43.12,57.40]$  & 2$\sigma$-1$\sigma$   \\ \hline
		\vspace{0.1cm}			LI ($\alpha > 0$) & $\log \alpha$ & \makecell{-2.52,-2.15,-1.77 \\-1.39, -1.02,-0.64 \\-0.27,0.11} & $[41.20,56.16]$   & 1$\sigma$  \\ \hline
		\vspace{0.1cm}			LI ($\alpha < 0$) & $\alpha$ & $[-0.29,-0.10]$  & \makecell{43,46,50 \\ 53,57,61}  & 2$\sigma$-1$\sigma$  \\
		\\
		\hline
	\end{tabular*}
\end{table*}

\begin{table*}
	\centering
	\caption{Results:  The first column details  the inflationary potential considered; the second column refers to the  potential's parameter; the third column reports the parameter  values for the standard inflationary model  that are not discarded by recent CMB and BAO data; the fourth column reports  the $\DN$ values for the standard inflationary model  that are not discarded by recent CMB and BAO data; the fifth column refers to the confidence level for which the values reported in columns 3,4 are in agreement with recent CMB and BAO data.
	}
	\label{tabla3}
	%
	\begin{tabular*}{1.0\textwidth}{c @{\extracolsep{\fill}}cccc}\hline
		\multicolumn{1}{|c}{\textbf{Inf. model}}&
		\multicolumn{1}{|c}{\textbf{Parameter}}&
		\multicolumn{1}{|c}{\textbf{Param. values}}&
		\multicolumn{1}{|c}{\textbf{$\Delta N_*$}}&
		\multicolumn{1}{|c|}{\textbf{Conf. region}}
		\\
		\hline
		\\
		\vspace{0.1cm}	HI & - & - & $[49.96,55.68]$  & 2$\sigma$-1$\sigma$ \\ \hline
		\vspace{0.1cm}		LFI & $p$ &  1,2   & \makecell{$[39.29,52.43]; 57.49$} & 2$\sigma$-1$\sigma$ \\ \hline
		\vspace{0.1cm}		RCQI ($\bar{\omega}_{\text{reh}} = 0$) & $\log \alpha$   & None & None   & None \\ \hline
		\vspace{0.1cm}		RCQI ($\bar{\omega}_{\text{reh}} = 1/3$) & $\log \alpha$   & None & None  & None \\ \hline
		RCHI & $A_I$ & \makecell{$[-0.29,23.86]$ \\  $[61,100]$}  & $[42.81,55.68]$  & 2$\sigma$-1$\sigma$ \\ \hline
		RCMI & $\log \alpha$ & -4.3,-4  & 55.21,56.65,56,71   & 2$\sigma$ \\ \hline
		\vspace{0.1cm}			NI & $\log (f/M_P)$ & 0.85,1  & $[53.02,57.20]$  & 2$\sigma$ \\ \hline
		\vspace{0.1cm}			ESI ($\bar{\omega}_{\text{reh}} = 0$ ) & $\log q$ & \makecell{-3,-1.3,-1,-0.6 \\ -0.3,0,0.18,0.54} & $[42.51,56.87]$  & 2$\sigma$-1$\sigma$ \\ \hline
		\vspace{0.1cm}			ESI ($\bar{\omega}_{\text{reh}} = -1/3$)  & $\log q$ & \makecell{-3,-1.3,-1,-0.6 \\ -0.3,0,0.18,0.54}& $[39.17,54.77]$  & 2$\sigma$-1$\sigma$ \\ \hline
		\vspace{0.1cm}			PLI & $\log \alpha$  & None & None   & None \\ \hline 
		\vspace{0.1cm}			DWI &  $\log (\phi_0/M_P)$ & $[1.18,3]$  & $[50,57.48]$   & 2$\sigma$ \\ \hline
		\vspace{0.1cm}			LI ($\alpha > 0$) & $\log \alpha$ & \makecell{-2.52,-2.15,-1.77 \\ -1.39, -1.02,-0.64 \\-0.27,0.11} & $[41.20,56.16]$ & 2$\sigma$-1$\sigma$ \\ \hline
		\vspace{0.1cm}			LI ($\alpha < 0$) & $\alpha$  & $[-0.35,-0.10]$  & \makecell{43,46,50 \\ 53,57,61}  & 2$\sigma$-1$\sigma$ \\
		\\
		\hline
	\end{tabular*}
\end{table*}

\section{Summary and Conclusions}\label{conclusions}

\label{conclusions}

In this paper we have analyzed the feasibility of a representative set of inflation potentials in the context of both the standard inflationary framework and the self-induced collapse of the inflaton's wave function proposal. For this, we have performed a statistical analysis using recent CMB and BAO data to obtain the confidence interval for $n_s$ in the context of a particular collapse model: the \textit{Wigner} scheme. Then, we have compared the predictions of each potential for $n_s$ given by \wig $ $ collapse scheme with the $2\sigma$ confidence interval resulting from the statistical analysis with observational data. The same analysis was also performed for the prediction of $n_s$ in the standard inflationary model, but in this case, the comparison was also performed considering  confidence regions in the $n_s-r$ plane. The reason for not including the latter  in the analysis of the collapse model is that these models predict a strong suppression of primordial tensor modes.

In \wig $ $ scheme,  the predicted scalar spectral index is given by $n_s = 1 +2\ei - \eii$, and the observational data suggests that $n_s \leq 1$ (at $2\sigma$ CL). Henceforth, any inflationary potential that satisfies $2\ei \leq \eii$ will be consistent with the data at 2$\sigma$ CL within \wig $ $ collapse scheme.  This result, together with the generically predicted smallness of $r$, relaxes the constraints on allowed  potentials with respect to the standard inflationary model. Specifically, within the collapse framework, constraints from observational data allow for inflationary potentials that are not as concave as the ones required by the standard model.
	
 In particular, the corresponding potentials of DWI, RCMI and NI (see Table \ref{listapotenciales}) are very well motivated models from the theoretical point of view and in good agreement with the data when considering \wig $ $ scheme (see Table \ref{tabla2}). On the other hand, in the standard scenario, those same potentials are barely consistent with the data (and possibly will be discarded if future data sets bound $r \leq 0.01$ see Table \ref{tabla3}). Notably, the three aforementioned potentials are not particularly as concave as the ones in HI, RCHI, ESI and LI (with $\alpha >0$), which are in perfect agreement with the data in the standard model and in \wig $ $ scheme.  In contrast, full convex potentials such as LFI for $p \geq 2$ and PLI are not favored by the data in any approach.

The fact that full concave potentials (instead of convex ones) are favored by the data in the standard scenario can be explained as follows. The data indicates that $n_s \leq 1 $, hence the contribution of $\eii$ should dominate in the predicted expression $n_s^\std = 1 -2\ei - \eii$ because $2\ei$ is required by the data to be as small as $r^\std$ (recall that $r^\std=16\ei$). From the expression of $\ei$ and $\eii$ in terms of the potential and its derivatives (see Eqs. \ref{PSR}), one can see that as observational bounds on $r$ decrease, the potentials' concavity should increase.  In \wig $ $ scheme concave potentials are  favored by the data as well but also potentials whose shape satisfies $2\ei \leq \eii$. That condition enlarges the families of potentials allowed, including the ones in the border between  convex and concave type.  In particular, the RCQI model is ruled out in the context of the standard framework while the prediction of the \textit{Wigner} collapse model is in agreement for a range of the free parameters.

We end our work by stressing that the difference in the predicted expressions for $n_s$ and $r$ with respect to the standard model is due to the self--induced collapse proposal and the semiclassical gravity assumption. This shows that when facing conceptual issues such as the quantum measurement problem in the early Universe, the possible solutions might lead to novel predictions that can be compared with observational data.

%
\section*{Acknowledgments}
The authors are supported by PIP 11220120100504 (CONICET) and grant G140 (UNLP).


\bibliography{bibliografia0}
\bibliographystyle{apsrev}

\end{document}